\pdfoutput=1
\documentclass[12pt]{iopart}

\usepackage{iopams}  

\usepackage{xcolor}
\usepackage{epsfig}
\usepackage{graphicx}
\usepackage{amsfonts}
\usepackage[figuresright]{rotating}
\usepackage{amssymb}
 \expandafter\let\csname equation*\endcsname\relax
  \expandafter\let\csname endequation*\endcsname\relax
\usepackage{amsmath}
\usepackage{color}

\usepackage{graphicx}
\DeclareGraphicsExtensions{.eps, .PNG, .jpg}

\newcommand{\ket}[1]{| #1 \rangle}
\newcommand{\bra}[1]{\langle #1 |}

\begin{document}

\title[]{Entanglement of a 3D generalization of the Kitaev model on the diamond lattice}

\author{Ian Mondragon-Shem and Taylor L. Hughes}

\address{Department of Physics, University of Illinois, Urbana-Champaign, 1110 West Green St, Urbana IL 61801}
\ead{mondrag2@illinois.edu}
\vspace{10pt}
\begin{indented}
\item[]\today
\end{indented}

\begin{abstract}
We study the entanglement properties of a three dimensional generalization of the Kitaev honeycomb model proposed by Ryu [Phys. Rev. B 79, 075124, (2009)]. The entanglement entropy in this model separates into a contribution from a $Z_2$ gauge field and that of a system of hopping Majorana fermions, similar to what occurs in the Kitaev model. This separation enables the systematic study of the entanglement of this 3D interacting bosonic model by using the tools of non-interacting fermions. In this way, we find that the topological entanglement entropy comes exclusively from the $Z_2$ gauge field, and that it is the same for all of the phases of the system. There are differences, however, in the entanglement spectrum of the Majorana fermions that distinguish between the topologically distinct phases of the model. We further point out that the effect of introducing vortex lines in the $Z_2$ gauge field will only change the entanglement contribution of the Majorana fermions. We evaluate this contribution to the entanglement which arises due to gapless Majorana modes that are trapped by the vortex lines. 
\end{abstract}

%
%
\submitto{\JSTAT}
%
\maketitle

\section{Introduction}

Over the past decade, the characterization and classification of topological insulator and superconductor (symmetry protected) phases of matter has played a central role in condensed matter research  \cite{Hasan2010,Qi2011}. These phases cannot be adiabatically connected with a trivial atomic limit, they typically exhibit gapless excitations at the sample boundary. Furthermore,  there are topologically ordered phases with long-range entanglement that go beyond the symmetry-breaking classification of states of matter. Topologically ordered states are a particular type of topological phase that arises in two or more dimensions and is characterized by a topology-dependent ground state degeneracy and long-range entanglement \cite{Wen2007,Wen1990}. 

The Kitaev honeycomb lattice model is a paradigm model  for the study of topologically ordered states \cite{Kitaev2006}. Its discovery was an important milestone because it is one of the first models that could be solved exactly which exhibits topological order and phase transitions between states with abelian and non-abelian excitations. Various extensions have been devised in two and three dimensions \cite{Wu2009, Ryu2012,Ryu2009,Chern2010}. In the three-dimensional case, long-range entanglement does not necessarily imply a topologically ordered state, although there can be contributions to long-range entanglement that are of topological origin \cite{Grover2011}, and point-like excitations always obey bosonic or fermionic statistics.  In spite of this, there  is nevertheless an interest in understanding generalizations of the Kitaev model to three dimensions because their study can lead to insights into the topological nature of certain interacting bosonic systems.

One fundamental way to characterize topological phases of matter is with spatial entanglement. Entanglement is an important tool used in condensed matter research to study  properties of the phases of a system and the phase transitions that separate such phases \cite{Amico2008,Vidal2003, Song2012,Chiara2012, Hsu2009,Li2008}. In particular, it has provided numerous insights related with both non-interacting topological insulators and superconductors \cite{Turner2010, Fidkowski2010,Pollman2010, Prodan2010, Hughes2011} as well as interacting topological phases \cite{KitaevPreskill2006, Thomale2010,Qi2012, Regnault2009, Flammia2009}. Recently, it was shown by Yao and Qi \cite{Yao2010}  that the entanglement of the 2D Kitaev model can be understood as arising from two contributions: one describing an emergent static $Z_2$ gauge field, and the second from non-interacting Majorana fermions hopping on a lattice. This insight revealed the origin of the topological entanglement entropy of the Kitaev model, and clarified the difference between its abelian and non-abelian phases in terms of its entanglement properties.

In this work we explore the extension of these results to three dimensions. We show that the same property of the entanglement found in the Kitaev model holds for a three-dimensional generalization proposed by Ryu \cite{Ryu2009}.  We explore the entanglement properties of this model in terms of signatures identifying the various phases of the system. We point out that introducing vortex defects in the $Z_2$ gauge field does not affect the factorization property of the density matrix, so that the entanglement contribution arising from these defects is determined by gapless Majorana degrees of freedom that are trapped by vortex configurations in the $Z_2$ gauge field. We show examples of the effect of such vortex lines on the entanglement of the system.

\section{Entanglement properties of the Kitaev model}

\subsection{Kitaev's honeycomb model}

As a warm-up, we review the 2D Kitaev model and its entanglement properties in this section. Consider a honeycomb lattice with a spin-$1/2$ degree of freedom represented by Pauli matrices $\sigma^a$ ($a=1,2,3$) at each lattice site.  Because of the geometry of the honeycomb lattice, each site has three nearest-neighbors. We label the three possible vectors connecting a lattice site to its  nearest-neighbors as $x$, $y,$ and $z$-links. The Kitaev model is obtained by assigning anisotropic exchange couplings between nearest neighboring spins according the type of link that connects them
\begin{equation}
H=-\sum_{x-\text{link}}J_x \sigma^x_i \sigma^x_j-\sum_{y-\text{link}}J_y \sigma^y_i \sigma^y_j-\sum_{z-\text{link}}J_z \sigma^z_i \sigma^z_j. 
\end{equation}
This particular form of exchange interaction makes this model exactly soluble. In particular, its eigenstates can be obtained explicitly by writing the spin degrees of freedom in terms of Majorana fermion operators. This is the method we will follow in this work, although one can also obtain the eigenstates through a Jordan-Wigner type transformation \cite{Feng2007}.

The main idea is to describe the two-dimensional Hilbert space of a spin-$1/2$ degree of freedom using a set of four Majorana fermions $\{b^{x}_i,b^{y}_i,b^{z}_i, c_i \}$ which are defined in an enlarged four-dimensional Hilbert space. These Majorana operators satisfy $b_i^2=1$, $c_i^2=1$, $\{b^{\alpha}_i,b^{\beta}_j\}=2\delta_{ij}\delta_{\alpha, \beta}$ and $b^{\alpha}_i c_j=-c_j b^{\alpha}_i$.  If one defines  $\tilde{\sigma}^\alpha_{i} = ib^{\alpha}_i c_i$, then this operator is a consistent representation of $\sigma^{\alpha}_i$ if we impose a constraint that restricts $\tilde{\sigma}_i^{\alpha}$ to a two-dimensional Hilbert space. This constraint is found by noting that the $\tilde{\sigma}^{\alpha}_i$ operators commute with the product $D_i=i b^x_i b^y_i b^z_i c_i$. Since $D^2_i=1$, we can impose the constraint $D_i=1$ to restrict $\tilde{\sigma}^{\alpha}_i$ to the desired two-dimensional Hilbert space. One can check that $\tilde{\sigma}^{\alpha}_i$ defined with this constraint satisfies the same algebra as the original spin operators. Hence, $\tilde{\sigma}_i^{\alpha}$ consistently describes the original spin-$1/2$ degree of freedom. 

In terms of these new operators, the Kitaev model takes the form
\begin{equation}
\tilde{H}=\frac{i}{2}\sum_{\langle j,k \rangle}J_{\alpha_{jk}}\hat{u}_{jk}c_j c_k,
\end{equation}
where $\hat{u}_{jk}=i b^{\alpha_{jk}}_j b^{\alpha_{jk}}_k$ are referred to as link operators, with $\alpha_{jk}=x,y,z$ depending on whether the $j$ and $k$ indices form a $x$, $y,$ or $z$-link. A consistent sign  convention is to choose the $j$ index to label a site in the $\mathcal{A}$ sublattice, and correspondingly $k$ in the  $\mathcal{B}$ sublattice. The fundamental advantage that is gained from using the Majorana fermion language is made apparent by noting that the link operators satisfy
\begin{equation}
\left[\tilde{H},\hat{u}_{jk}\right]=0 \quad \text{and} \quad \left[\hat{u}_{jk},\hat{u}_{lm}\right]=0.
\end{equation}
We can thus diagonalize the Hamiltonian and the $\hat{u}_{jk}$ operators simultaneously.  An eigenstate of the Kitaev model can then be labeled by a fixed configuration of eigenvalues of the link operators. Since $\hat{u}^2_{jk}=1$, the link operators  only have two eigenvalues $\pm1$. Once a configuration of eigenvalues is chosen, what remains is a Hamiltonian of free Majorana fermions $c_i$ hopping on a lattice, which can be solved straightforwardly. The effect of the link operators will be at most to change the signs of the hopping elements. This means the link operators effectively act like a static $Z_2$ gauge field that couples to the $c_i$ Majorana fermions. 

To obtain the ground state, we need to know what configuration of the $Z_2$ gauge field leads to the lowest overall energy.  Lieb showed that this configuration corresponds to all $u_{jk}=1$ \cite{Lieb1994}. Using this configuration, and solving for the corresponding Majorana fermion ground state $\ket{\phi(u)}$, one can then calculate the physical state $\ket{\Psi}$ by projecting into the sector in which $D_i=1$ for all $i$. This amounts to averaging over all possible gauge transformations of the $Z_2$ gauge field:
\begin{equation}
\ket{\Psi}=\frac{1}{\sqrt{2^{N+1}}}\sum_g D_g \ket{u}\otimes \ket{\phi(u)}.
\end{equation}
Here, $N$ is the total number of sites, $D_g=\prod_{i \in g} D_i$ with $g$ being a subset of lattice sites, and the sum runs over all possible subsets of sites. Other energy eigenstates can be obtained by the same gauge averaging procedure with some initial configuration of $Z_2$ fluxes.

Since the ground state has constant phases on the links, the Hamiltonian in this sector is translationally invariant. A change of basis to momentum space leads to the following two-by-two single-particle Hamiltonian
\begin{equation}
h(\mathbf{k})=-\text{Re}\phi(\mathbf{k}) \tau^y-\text{Im}\phi(\mathbf{k}) \tau^x,
\end{equation}
where $\tau^{a}$ ($a=0,x,y,z$) are Pauli matrices that act on the sublattice index, and $\phi(\mathbf{k})=J_x e^{i\mathbf{k}\cdot \mathbf{a}_1}+J_y e^{i\mathbf{k}\cdot \mathbf{a}_2}+J_z$ with $\mathbf{a}_{1,2}$ the primitive vectors that generate the $\mathcal{A}$ hexagonal sublattice. The energy spectrum of this Hamiltonian is $\epsilon_{\pm}(\mathbf{k})=\pm \vert\phi(\mathbf{k})\vert$. 

Due to the form of this spectrum, one can divide the space of parameters into two regions. Whenever the couplings satisfy the inequalities $\vert J_x \vert\le \vert J_y \vert+\vert J_z \vert$, $\vert J_y \vert\le \vert J_z \vert+\vert J_x \vert,$ and $\vert J_z \vert\le \vert J_x \vert+\vert J_y \vert$, the spectrum is gapless due to the time-reversal invariance of the Majorana fermion Hamiltonian. The spectrum in this parameter regime can thus be gapped out by the addition of three-spin interactions that break time-reversal symmetry. The resulting gapped ground state has quasiparticle excitations that obey non-abelian statistics, and so this phase is referred to as the non-abelian phase of the Kitaev model. By contrast, if the triangular inequalities of the $J_\alpha$ are not satisfied then the system is gapped without the need of any additional terms. In this case the excitations satisfy abelian statistics, and so in this case the system realizes an abelian phase. 

\subsection{Entanglement of quantum states}

Let us now briefly review how to quantify the entanglement of a state $\ket{\Omega}$. One starts by choosing a partition of the Hilbert space into two complementary subspaces, say $A$ and $B$. The entanglement between these two parts of the Hilbert space can be quantified by the so-called von-Neumann entropy, defined as
\begin{equation}
S_A=-\text{Tr}_A\left(\rho_A \log \rho_A\right).\label{SA}
\end{equation}
The reduced density matrix of region $A$ is given by $\rho_A=\text{Tr}_B\left[\ket{\Omega}\bra{\Omega}\right]$, where $\text{Tr}_B$ denotes the trace over the degrees of freedom in $B$. We will refer to the partitioning of the Hilbert space as an entanglement cut that is performed on the system. A generalization of the entanglement entropy that has also been useful in characterizing condensed matter systems, namely the Renyi entropy, is given by
\begin{equation}
S^{(n)}_A=\frac{1}{1-n} \text{Tr}_A\left[\rho_A^n\right].
\end{equation}
We can recover the von Neumann entropy by taking the limit $S_A =\lim_{n\rightarrow 1} S^{(n)}_A$. This form of the entanglement entropy in terms of the quantity $\text{Tr}_A\left[\rho_A^n\right]$ will be useful for calculating the entanglement of the Kitaev model and its 3D generalization.

As we will discuss in the following sections, the entanglement spectrum of Kitaev-type models can be reduced to the computation of the entanglement of quadratic fermionic Hamiltonians.  In such cases, the entanglement entropy is completely determined by the eigenvalues $\{\zeta_i\}$ of the correlation matrix $\left[C\right]_{ij}=\bra{\Omega}c^{\dagger}_i c_j\ket{\Omega}$\cite{peschel2003,peschel2009}, where the $i,j$ indices are restricted to the $A$ subspace. The entanglement entropy $S_A$ in terms of this set of eigenvalues is then given by
\begin{equation}
S=\sum_i \left\{ -\zeta_i \ln \zeta_i-(1-\zeta_i)\ln(1-\zeta_i)\right\}.\label{Sspect}
\end{equation}
The set $\{\zeta_i\}$ is called the single-particle entanglement spectrum and it corresponds to the eigenvalues of the correlation matrix. The entanglement entropy, and additionally, all entanglement quantities of  a free-fermion ground state $\ket{\Omega}$ can thus be understood by analyzing the $\zeta_i.$  The $\zeta_i$  lie between $0$ and $1,$ and thus the closer the modes are to $1/2,$ the larger the entanglement of a subsystem.  The distribution of the $\zeta_i$ is what we will keep track of in the discussion that follows.

\subsection{Entanglement in Kitaev's honeycomb model}

The phases of the Kitaev model were characterized in \cite{Yao2010} using entanglement. In general, computing the entanglement of an interacting spin model can be challenging both analytically and numerically. However, because of the special structure of the Kitaev model, this task is dramatically simplified. The entanglement of an eigenstate $\ket{\psi}$ of the Kitaev model can be obtained by separately calculating the entanglement of the $Z_2$ gauge field and the Majorana fermions. More specifically, Yao and Qi showed that the following relation holds
\begin{equation}
\text{Tr}_A\left[\rho_{A}^n\right]=\text{Tr}_{A,G}\left[\rho_{A,G}^n\right]\cdot \text{Tr}_{A,F}\left[\rho_{A,F}^n\right]
\end{equation}
where $\rho_{A}=\text{Tr}_B \left[\ket{\psi}\bra{\psi}\right]$, the reduced density matrix $\rho_{A,F}$ ($\rho_{A,G}$) describes the Majorana fermions (a pure $Z_2$ gauge field) in region $A$, and the trace $\text{Tr}_{A,F(G)}$ runs over the fermion (gauge) degrees of freedom in region $A$. The factorization of $\text{Tr}_A\left[\rho_{A}^n\right]$ is useful because, by taking the limit $n \rightarrow 1$, one finds that the entanglement entropy can be written as 
\begin{equation}
S_{A}=S_{A,G}+S_{A,F},
\end{equation}
where $S_{A,F(G)}$ is the entanglement entropy of the fermions (gauge field). 

Using this insight, Yao and Qi found that in both abelian and non-abelian phases the entanglement entropy of the Kitaev model can generically be written as $S_A=\left(\alpha +\log 2\right)L-\log 2$, where $\alpha$ is a non-universal constant and $L$ is the length of the boundary separating regions $A$ and $B$. The term proportional to $L$ is the well-known area (perimeter in 2D) law for gapped states. The term that is independent of the boundary size is thus identified as the topological entanglement entropy. It is the same for all phases of the Kitaev model, and it arises exclusively due to the presence of the $Z_2$ gauge field.

It was further argued that, although the topological entanglement entropy is the same for both abelian and non-abelian phases, there is nevertheless a way in which their entanglement properties can distinguish these phases. Specifically, in the non-abelian phase the Majorana fermion ground state acquires a nonzero Chern number that leads to the presence of gapless states at the boundary. The presence of these boundary states leads to spectral flow in the entanglement spectrum and further contributes to the entanglement of the system. Such edge states do not generically arise in the abelian phase, so there are no additional entanglement contributions in this phase. This distinction was argued by Yao and Qi to be related to the nature of the quasiparticles in the non-abelian phase. Hence, they argued, the intrinsic difference between the abelian and non-abelian phases is manifested in their entanglement properties.

Although we will not make connections to the statistics of excitations in Ryu's 3D model, we will nevertheless find analogous behavior concerning the entanglement properties of its eigenstates. In particular, the entanglement entropy is also separable into gauge field and Majorana fermion components, and this insight allows one to understand and distinguish the topological phases of the system depending on the surface states (or absence thereof) in each phase, as we will see in later sections.

\begin{figure}
\begin{center}
\includegraphics[trim =0cm 2cm 0cm 0cm,scale=0.4]{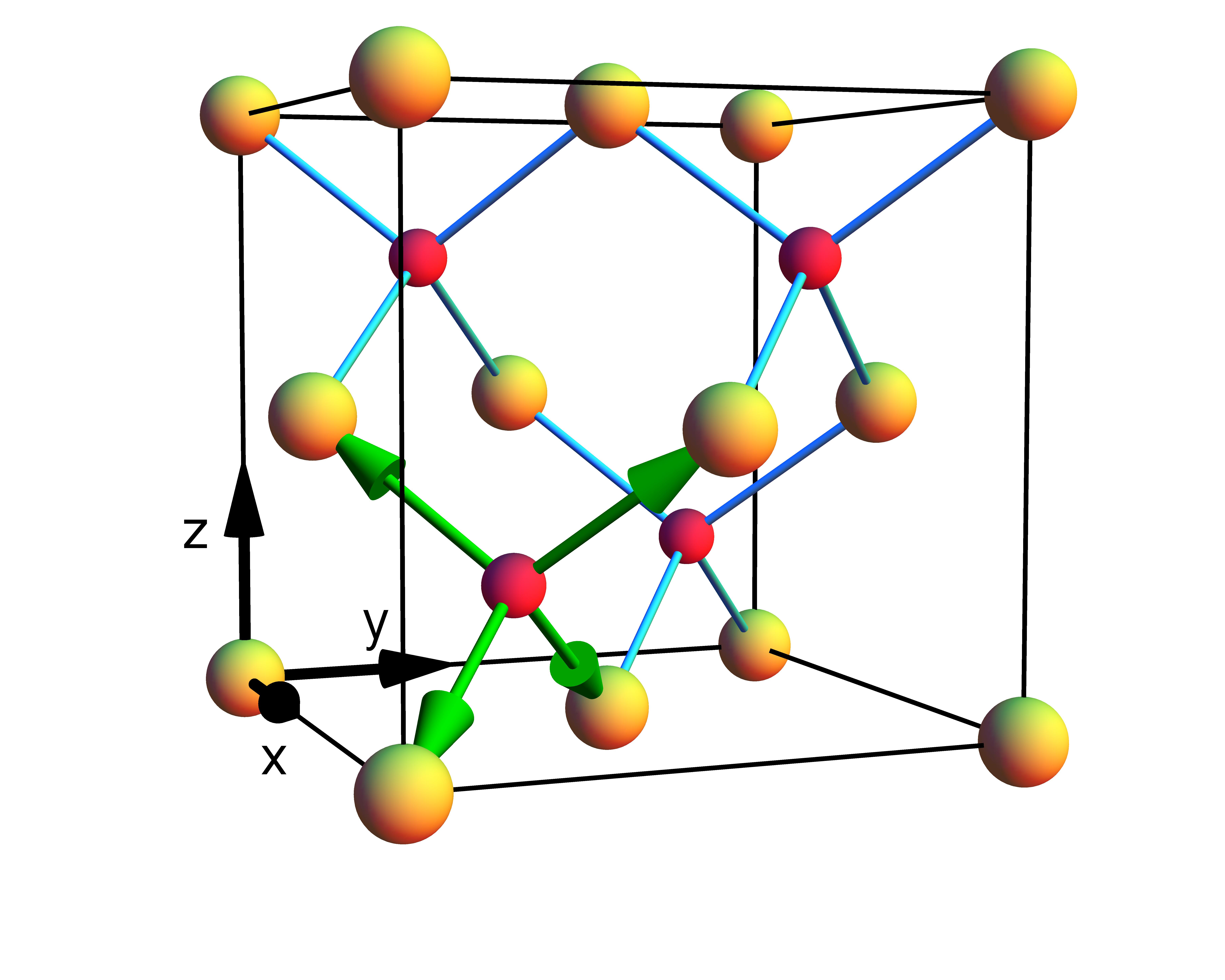}
\caption{Conventional cell of the fcc lattice. The yellow large (red small) spheres correspond to the $\mathcal{A}$ ($\mathcal{B}$) sublattice. The green arrows emanating from one of the $\mathcal{B}$ sites correspond to the $\mathbf{s}_i$ vectors. Corresponding to each of the bars connecting an $\mathcal{A}$ and $\mathcal{B}$ site there is a link operator $\hat{u}_{r_A r_B}$ with a specific value of the $Z_2$ gauge field.}\label{lattice}
\end{center}
\end{figure}

\section{Generalization of the Kitaev model to 3D}

In this section we discuss Ryu's model and its entanglement properties \cite{Ryu2009};  we will refer it as the Ryu-Kitaev diamond (RKD) model. The overall structure of the RKD Hamiltonian is analogous to that of the Kitaev honeycomb model, namely one considers anisotropic exchange couplings between nearest-neighboring spin degrees of freedom. Upon introducing Majorana operators, the Hamiltonian reduces to a problem of free Majorana fermions hopping in the presence of a $Z_2$ gauge field; in this case in three dimensions. There is, however, a fundamental difference with respect to the Kitaev model, namely the RKD model is designed to preserve time-reversal symmetry for all of its phases. This will have an important impact on the topological classification of the ground state which is manifested in its entanglement properties. 

\subsection{Hamiltonian and Majorana fermion description}

The RKD model is realized on the diamond lattice. The diamond lattice is formed by two fcc sublattices $\mathcal{A}$ and $\mathcal{B}$ that are shifted by a vector $\frac{a}{4}(-1\,1\,-1)^T$ ($a$ is the lattice constant of the fcc conventional cell). We choose the primitive vectors $\mathbf{a}_1=\frac{a}{2}\left(1\, 1\, 0 \right)^T$, $\mathbf{a}_2=\frac{a}{2}\left(0\, 1\, -1 \right)^T$ and $\mathbf{a}_3=\frac{a}{2}\left(1\, 0\, -1 \right)^T$. Each lattice site in the diamond lattice has four nearest neighbours, and the vectors connecting a site in sublattice $\mathcal{A}$ to its nearest neighbours are: $\mathbf{s}_1=\frac{a}{4}\left(1\, 1\, \,1 \right)^T$, $\mathbf{s}_2=\frac{a}{4}\left(-1\, -1\, 1\right)^T$, $ \mathbf{s}_3=\frac{a}{4}\left(1\,1\,-1\right)^T$, $ \mathbf{s}_0=\frac{a}{4}\left(-1\,1\,-1\right)^T$. We illustrate the structure of the diamond lattice together with the $\mathbf{s}_i$ vectors in Fig.\ref{lattice}. 

On each lattice site we place two spin-$1/2$ degrees of freedom $\sigma^a$ and $\tau^a$ ($a=0,1,2,3$). For convenience, we define $\alpha^{1,2,3}_j=\sigma^{1,2,3}_j \tau^x_j,$ $\alpha^0_j=\sigma^0_j \tau^z_j,$ $\zeta^{1,2,3}_j=\sigma^{1,2,3}_j \tau^z_j,$ and $\zeta^0_j=\sigma^0_j \tau^z_j.$  We then couple nearest-neighboring pairs of spins in the following anisotropic way
\begin{equation}
H=-\sum_{\mu=0}^{3}  \sum_{\mu-\text{links}} J_\mu\left(\alpha_j^\mu\alpha_k^\mu+\zeta_j^\mu\zeta_k^\mu\right).
\end{equation}
Here, the values $\mu=0,\ldots, 4$ label the four possible nearest neighbors determined by the $\mathbf{s}_\mu$. The key feature of this model is again the anisotropic nature of the exchange interactions. Similar to the Kitaev model, the eigenstates of the Hamiltonian are obtained by introducing Majorana degrees of freedom at each site. In the present case, since there is a four-dimensional Hilbert space at each site, we can consider an enlarged eight-dimensional Hilbert space with six Majorana fermions $\lambda^p_i$ ($p=0,\ldots,5$). The eigenstates are constrained to be in the subspace where $D_i=i\prod_{p=0}^5 \lambda_i^p=1$. By making the identification $\alpha_i^{\mu}=i\lambda_i^{\mu}\lambda_i^4$ and $\zeta_i^{\mu}=i\lambda_i^{\mu}\lambda_i^5$, the Hamiltonian becomes
\begin{equation}
H=i\sum_{\mu=0}^{3} J_\mu \sum_{\mu-\text{links}} \hat{u}_{jk}\left(\lambda_j^4\lambda_k^4+\lambda_j^5\lambda_k^5\right).
\end{equation}
where the link operators are given by $\hat{u}_{jk}=i \lambda^{\mu_{jk}}_j \lambda^{\mu_{jk}}_k$. Here, the link operators are again defined to go from sublattice $\mathcal{A}$ to sublattice $\mathcal{B}$. These link operators commute with the Hamiltonian, so we can replace them by a specific choice of eigenvalues. What remains is then a hopping model of two flavors of Majorana fermions that feel the same $Z_2$ field.  The RKD model includes additional interactions between spins on three neighboring sites which are introduced in order to remove non-generic degeneracies in the energy spectrum. This effectively leads to the following second-nearest neighbor hoppings in the Majorana fermion language:
\begin{eqnarray}
H_z&=&\sum_{r_A}\left[iK^z\left(\hat{u}_{r_A\,r_A-s_1}\hat{u}_{r_A\, r_A-s_3}\right)\lambda^T_{r_A-s_1} s^z \lambda_{r_A-s_3}\right] \nonumber\\
&+&\sum_{r_B}\left[iK^z\left(\hat{u}_{r_B+s_1\, r_B}\hat{u}_{r_B+s_3\,r_B}\right)\lambda^T_{r_B+s_1} s^z \lambda_{r_B+s_3}\right],\nonumber\\
H_x&=&\sum_{(i,j)\in \Lambda}\left\{\sum_{r_A}\left[iK^x\left(\hat{u}_{r_A\,r_A-s_i}\hat{u}_{r_A\, r_A-s_j}\right)\lambda^T_{r_A-s_i} s^x \lambda_{r_A-s_j}\right] \right. \nonumber\\
&+&\left.\sum_{r_B}\left[iK^x\left(\hat{u}_{r_B+s_i\, r_B}\hat{u}_{r_B+s_j\,r_B}\right)\lambda^T_{r_B+s_i} s^x \lambda_{r_B+s_j}\right]\right\},\nonumber
\end{eqnarray}
where the Pauli matrices $s^a$ ($a=0,x,y,z$) act on the $4,5$ indices and $\lambda^T=(\lambda^4, \,\, \lambda^5)$. Note the distinction of indices for the lattice vectors ${\bf{s}}_a$ and the Pauli matrices $s^a.$ The pair of indices $(i,j)$ runs over the set $\Lambda=\{(0,2),\,(2,3),\, (3,0)\}$.

Since the plaquettes of the diamond lattice are also hexagons, the ground state continues to occur when $\hat{u}_{jk}=1$ for all $j,k$, so the ground state is translationally invariant. With periodic boundary conditions,  the single-particle momentum space Bloch Hamiltonian is
\begin{eqnarray}
h(\mathbf{k})&=&\Theta^x(\mathbf{k})c^z s^x+\Theta^z(\mathbf{k}) c^z s^z-\text{Re}\Phi(\mathbf{k}) c^y s^0-\text{Im}\Phi(\mathbf{k}) c^x s^0,
\end{eqnarray}
where $c^a$ ($a=0,x,y,z$) are additional Pauli matrices acting on the sublattice degree of freedom, and we defined the functions
\begin{eqnarray}
\Phi(\mathbf{k})&=&J_0 e^{i\mathbf{k}\cdot \mathbf{a}_2}+J_1 e^{i\mathbf{k}\cdot \mathbf{a}_1}+J_2+J_3 e^{i\mathbf{k}\cdot \mathbf{a}_3},\\
\Theta^x(\mathbf{k})&=&K^x\sum_{(i,j)\in \Lambda}\sin \mathbf{k}\cdot (\mathbf{s}_i-\mathbf{s}_j),\\
\Theta^z(\mathbf{k})&=&K^z \sin  \mathbf{k}\cdot (\mathbf{s}_1-\mathbf{s}_3). 
\end{eqnarray}
The energy spectrum of the single-particle Hamiltonian is given by $\epsilon_{\pm}(\mathbf{k})=\pm\sqrt{\vert \Phi\vert^2+\Theta^{x 2}+\Theta^{z 2}}$, where $\vert \Phi\vert^2=(\text{Re}\Phi)^2+(\text{Im}\Phi)^2$. By evaluating this spectrum for various values of the parameters, one finds that there are several distinct gapped phases separated by gapless critical points. These gapless points correspond to phase transitions between topologically distinct phases. In \cite{Ryu2009}, Ryu identified two main phases, namely a strong and weak topological phase.  We will discuss these phases and their entanglement in the following sections.

\subsection{Symmetries and topological phases}

Similar to the Kitaev model, the RKD has topologically distinct phases depending on the relative strengths of the hopping parameters. Let us consider the single-particle Majorana fermion Hamiltonian in momentum space. The Hamiltonian $h(\mathbf{k})$ satisfies particle-hole symmetry  $C h(-\mathbf{k})C^{-1}=-h(\mathbf{k}) $ ($C=\mathcal{K}$), and time-reversal symmetry $T h(-\mathbf{k}) T^{-1}=h(\mathbf{k})$ ($\mathcal{T}=i c^z s^y \mathcal{K}$), where $\mathcal{K}$ represents complex conjugation. Note that the time-reversal symmetry operator satisfies $T^2=-1$. Hence, the model we are considering belongs to the symmetry class DIII of the Altland-Zirnbauer classification of non-interacting fermions \cite{Altland1997}. Similar to 3D time-reversal invariant topological insulators \cite{Fu2007}, there are strong and weak topological states that can be obtained in this model as  discussed in \cite{Ryu2009}. 

The strong topological phase has robust gapless states on any surface that separates the bulk from the vacuum. It can be characterized by a $\mathbb{Z}$ topological invariant defined for 3D systems that satisfy chiral symmetry $SH+HS=0$, where $S$ is the chiral operator. In the present case, this operator corresponds to $S=c^z s^y$.  In the basis in which $S$ is diagonal, the operator $Q(\mathbf{k})=2P(\mathbf{k})-1$ can be written in block off-diagonal form (where $P(\mathbf{k})$ is the projection operator into the occupied states). Let us then define the matrix in the block-off diagonal of $Q(\mathbf{k})$ as $q(\mathbf{k})$. Then the integer-valued topological invariant is given by \cite{Schnyder2008}
\begin{equation}
\nu_{3D}=\int_{BZ}\frac{d^3 k}{24 \pi^2}\epsilon^{\mu \nu \rho}\text{tr}\left[\left(q^{-1}\partial_{\mu}q\right)\left(q^{-1}\partial_{\nu}q\right)\left(q^{-1}\partial_{\rho}q\right)\right],\label{nu3d}
\end{equation}
where for the RKD model the $q(\mathbf{k})$ matrix is
\begin{equation}
q(\mathbf{k})=\frac{1}{\epsilon_+(\mathbf{k})}\left(
\begin{array}{cc}
 i \Theta_x+\Theta_z & -\text{Im}\Phi -i\text{Re}\Phi\\
-\text{Im}\Phi +i\text{Re}\Phi & -i \Theta_x-\Theta_z
\end{array}
\right).
\end{equation}
As was verified in \cite{Ryu2009}, there is a parameter regime for which $\nu_{3D}\ne 0$, signaling a nontrivial 3D topological ground state. We will discuss in the next section particular realizations of the parameters for which $\nu_{3D}=\pm 1,$ and analyze the corresponding entanglement properties.

As to the weak topological states, these arise when some of the hopping parameters are reduced sufficiently so that the ground state is adiabatically connected to either decoupled topological layers or decoupled topological wires. In these phases, the system has boundary modes only on certain surfaces of the system depending on the direction of the layers or wires that realize the topological state. These boundary states are protected by translation symmetry and can be gapped out by introducing disorder that respects the symmetries of class DIII. As such, this phase is not robust, at least not in the same sense that the strong topological phase is robust. It has been argued, however, that if the disorder is respected on average, such boundary states can still survive \cite{Fu2012, Fulga2014}.

To be concrete, suppose $J_3,K_z$ are sufficiently smaller than the other couplings so that the ground state is adiabatically connected to the $K_z=J_3=0$ limit. In this case, the system can be viewed as a set of weakly coupled layers that are perpendicular to $\mathbf{s}_1$.  When the layers are completely decoupled, each layer realizes a two-dimensional system in class DIII,  which means that the ground state is classified by a $\mathbb{Z}_2$ invariant.  Because of this, the topology of the ground state of these layers is different from that of the Kitaev model, although similar to the square lattice model studied in \cite{Ryu2012}.

The $\mathbb{Z}_2$ invariant in class DIII is given by the Fu-Kane formula \cite{Fu2006,Ryu2012}
\begin{equation}
\nu_{2D}= \prod_{q: \text{TRIM}} \frac{\sqrt{\text{det}(w(q))}}{\text{Pf}(w(q))},\label{nu2d}
\end{equation}
where TRIM stands for the set of four time-reversal invariant momenta in the first Brillouin zone (FBZ) of the hexagonal lattice, $w_{nm}(\mathbf{k})=\bra{u_{n}(-\mathbf{k})}T\ket{u_{m}(\mathbf{k})}$, and $\text{Pf}[w]$ is the Pfaffian of the matrix $w(\mathbf{k})$. In \ref{2dnu}, we show the derivation for obtaining the following expression of this topological invariant
\begin{equation}
\nu_{2D}=\text{sign}(J_0+J_2+J_3) \text{sign}(-J_0+J_2-J_3)\text{sign}(-J_0+J_2+J_3) \text{sign}(J_0+J_2-J_3).\nonumber
\end{equation}
Using this expression, we find that there are parameter regimes for which $\nu_{2D}=-1$, indicating the presence of a nontrivial phase for each layer. 

If we now consider the case when another coupling, say $J_0$ is sufficiently small, then the system will be adiabatically connected to decoupled wires in class DIII. This class also has a $\mathbb{Z}_2$ classification. Following \cite{Ryu2012}, we again characterize the topological state by the Fu-Kane formula Eq. \ref{nu2d}. The main difference with the previous calculation is that now there are only two time-reversal invariant momenta. The calculation leads to 
\begin{equation}
\nu_{1D}=\text{sign}(J_2+J_3)\text{sign}(J_2-J_3).\label{nu1d}
\end{equation}
When $J_3$ is greater than $J_2$, this expression gives $\nu_{1D}=-1$. This leads to localized boundary modes for each wire. Upon coupling the wires to form the 3D bulk system, these boundary modes become dispersive and are generically susceptible to being gapped by disorder. The main difference with the case of weakly-coupled layers is that here there will be no spectral flow between the positive and negative energy bands, whereas in the layer case there is. This difference will manifest itself in the entanglement spectrum as we will see when we discuss the entanglement of the RKD model.

\section{Entanglement of the RKD model}

\subsection{Factorization of the trace of the density matrix}

In order to calculate the entanglement of the RKD model, we first show that the factorization found in two dimensions for the Kitaev model also holds for the RKD model. The derivation we present here is essentially an extension of the derivation by Yao and Qi, the main difference being that there are more Majorana operators per lattice site in the RKD model. In this section we provide a general description of how the derivation works, and we leave the details for \ref{proof}.

We start by writing the explicit form of an eigenstate of the RKD model. This eigenstate will be a product of the state of the $Z_2$ gauge field $\ket{u}$ and the corresponding Majorana fermion state $\ket{\phi(u)}$. By projecting into the $D_j=1$ subspace we obtain the physical state:
\begin{equation}
\ket{\psi}= \sqrt{\frac{1}{2^{1-N}}}\prod_{j}\left(\frac{1+D_j}{2}\right)\ket{u}\otimes \ket{\phi(u)}, \label{proj}
\end{equation}
where the product runs over all of the $N$ lattice sites of the system. The objective will be to calculate $\text{Tr}_A\left[\rho_A^n\right]$ in terms of the reduced density matrix of a pure $\mathbb{Z}_2$ gauge field $\rho_{A,G}$ and the reduced density matrix of the free Majorana fermions $\rho_{A,F}=\text{Tr}_B\left[\ket{\phi(u)}\bra{\phi(u)}\right]$. The main complication for achieving this is that the state $\ket{u}\otimes \ket{\phi(u)} $ is multiplied by the $D_j$ operators. Thus, it would seem that the $D_j$ operators will inevitably appear in the final expression upon taking the trace of powers of the density matrix. 

This can be resolved by explicitly performing the traces of the $Z_2$ gauge field over region $B$. This is achieved by rewriting the link operators that cross the entanglement cut in terms of new link operators that exist exclusively on either the $A$ or $B$ region. Upon taking the trace over region $B,$ and for each power $\rho_A^n$ that is computed, there will appear matrix elements of the operators $\{\lambda^0_i,\lambda^1_i,\lambda^2_i,\lambda^3_i\}$ which can be simplified explicitly.  After carrying out this procedure, the $\{\lambda^0_i,\lambda^1_i,\lambda^2_i,\lambda^3_i\}$ operators drop out of the expression. What remains at this stage are the $\{\lambda^4_i,\lambda^5_i\}$ operators which act on the fermion state $\ket{\phi(u)}$.

Because of the manner in which matrix elements of the $Z_2$ gauge field are traced out, it turns out that all of the $\{\lambda^4_i,\lambda^5_i\}$ can be arranged into operators that project into definite sectors of fixed fermion parity. By using the fact that the fermion parity of the fermion ground state is fixed, these fermion parity projectors can be simplified appropriately, to the point where there will no longer be any of the $\{\lambda^4_i,\lambda^5_i\}$ operators in the expression. The resulting expression turns out to be (see Appendix B)
\begin{equation}
\text{Tr}_A\left[\rho^n_A\right]=\frac{1}{2^{(n-1)(L-1)}} \text{Tr}_A\left[\rho_{A,F}^{n}\right],
\end{equation}
where $L$ is the number of links crossing the entanglement cut. By further noting that
\begin{equation}
\text{Tr}_{A,G}\left[\rho^n_{A,G}\right]=\frac{1}{2^{(n-1)(L-1)}},
\end{equation}
for a pure $Z_2$ gauge field, one then obtains the desired result
\begin{equation}
\text{Tr}_A\left[\rho_A^n\right]= \text{Tr}_{A,G}\left[\rho^n_{A,G}\right]\text{Tr}_{A,F}\left[\rho^n_{A,F}\right].
\end{equation}
Using this property of the density matrix of the RKD model, one can then proceed to calculate the entanglement of the system in the same way that it was done for the Kitaev model on the honeycomb lattice.

\begin{figure*}
\begin{center}
\includegraphics[trim =0cm 3cm 0cm 0cm,scale=0.28]{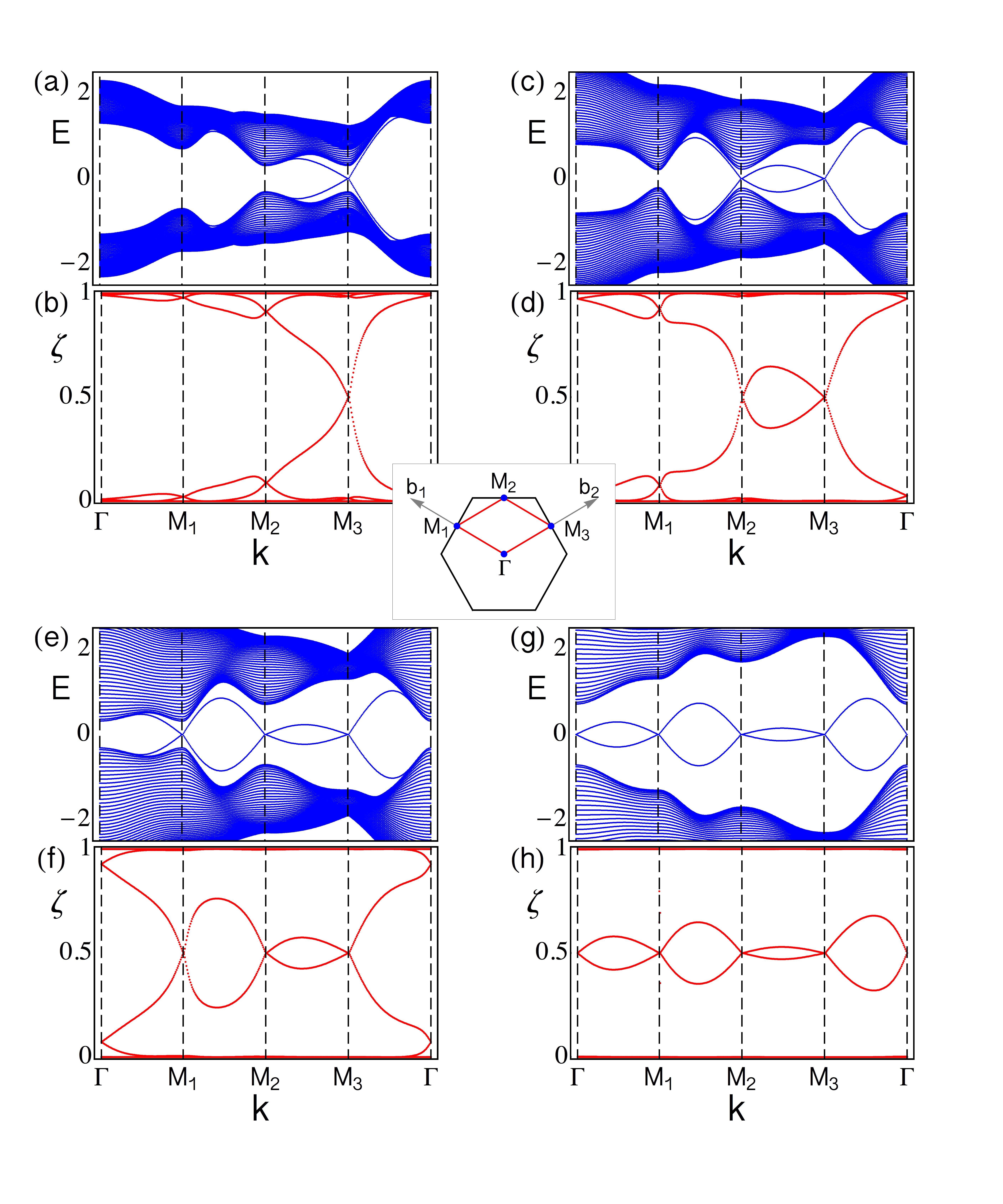}
\caption{Energy and entanglement spectrum in momentum space for $J_0=1$, $J_1=0.3$, $J_2=K_x=K_z=0.5$ and varying $J_3.$ We have $J_3=0.5$ (a,b), $J_3=1.0$ (c,d), $J_3=1.5$(e,f) and $J_3=2.5$. The center figure shows the Brillouin zone and the corresponding path in momentum space over which we evaluate both the energy and entanglement spectrum. The vectors $\mathbf{b}_1$ and $\mathbf{b}_2$ are reciprocal lattice vectors satisfying $\mathbf{b}_i\cdot \mathbf{a}_j=2\pi \delta_{ij}$, with $i,j=1,2$. The points $\Gamma$, $M_1$, $M_2$ and $M_3$ label the time-reversal invariant momenta of the hexagonal lattice generated by $\mathbf{a}_{1}$ and $\mathbf{a}_{2}$.}\label{spectra}
\end{center}
\end{figure*}

\subsection{Entanglement properties}

An immediate consequence of the factorization property of $\text{Tr}_A\left[\rho^n_A\right]$ is that there is a contribution to the entanglement entropy which does not scale with system size. This contribution arises exclusively from the $Z_2$ gauge field part and it is given by
\begin{equation}
S_{A,G}=\left(\log 2\right)L-\gamma_{\text{top}},
\end{equation}
where $\gamma_{\text{top}}=\log 2$ is the topological contribution to the entanglement entropy. In fact, this value of the topological entanglement entropy is the same as that of the two-dimensional $Z_2$ gauge field of the Kitaev model. This is consistent with the discussion in \cite{Grover2011}, where it was shown that the entanglement entropy of a discrete gauge theory of symmetry group $G$ would have a topological entanglement entropy $\gamma_{\text{top}}=\log \vert G \vert$ in both two and three dimensions, where $\vert G \vert$ is the number of elements in the group.

Let us now evaluate the entanglement of the Majorana fermion part of the ground state. Throughout, we will consider an entanglement cut that partitions the system along the plane generated by $\mathbf{a}_1$ and $\mathbf{a}_2$. We will maintain periodic boundary conditions along theses two directions. Since the ground state is translationally invariant, we can Fourier transform both directions and consider Hamiltonians that are dependent on the momenta $(\mathbf{k}_1, \mathbf{k}_2)$. The corresponding FBZ is depicted in the inset at the center of Fig.\ref{spectra}. This figure of the Brillouin zone also shows the path along which we will evaluate the energy and entanglement spectrum. The path includes the time-reversal momenta which are the momenta where the gap closing points occur in this model.

Even though the parameter space is significantly large, it will be sufficient to restrict ourselves to a specific set of parameters that will allow us to explore the relevant phases of the model. We thus fix the parameters $J_0=1$, $J_1=0.3$, $J_2=K_x=K_z=0.5$ and vary $J_3$. In Figs. \ref{spectra} a,c,e,g, we show the energy spectrum with open boundary conditions in the $\mathbf{a}_3$ direction. Each subfigure corresponds to four values of the parameter $J_3=0.5, 1.0,1.5, 2.5$, respectively. The corresponding entanglement spectrum is shown in Figs. \ref{spectra} b,d,f,h with periodic boundary conditions in the $\mathbf{a}_3$ direction.

The gap of the model closes as we continuously change between these values of $J_3$. Each time the gap closes, the system undergoes a topological phase transition. The four cases we show here thus correspond to four topologically distinct phases. For all four phases there are energy modes in the gap that cross zero energy. These are the surface states, which signal the nontrivial nature of the ground state. Correspondingly, the entanglement spectrum shows entanglement modes between $0$ and $1$ that behave in a similar way as the surface states. This is due to the fact that the correlation matrix is directly related with the spectrally flattened version of the single-particle Hamiltonian. Thus, the topological surface states will be manifest in the entanglement spectrum as entanglement modes that cross $1/2$ \cite{Turner2010}.

For $J_3=0.5,1.0,1.5$ there is spectral flow in both the energy and entanglement spectrum. Both the $J_3=0.5$ and $J_3=1.5$ phases correspond to strong topological states characterized by $\nu_{3D}=-1$ and $\nu_{3D}=1$ respectively, which we verified numerically using Eq. \ref{nu3d}. There is a single crossing in the $J_3=0.5$ case, whereas there are three crossings for $J_3=1.5$. In the intermediate case, namely $J_3=1.0$, we find that $\nu_{3D}=0$. By varying continuously the couplings $J_1$ and $K_z$ to zero, we have found no additional gap closings at the time-reversal invariant momenta, which means that this phase is adiabatically connected with the system of decoupled layers perpendicular to $\mathbf{s}_1$ we discussed earlier. By using Eq. \ref{nu2d}, we obtain $\nu_{2D}=-1$ in this phase, which confirms that the phase at $J_3=1.0$ corresponds to a weak topological phase of coupled 2D topological states in class DIII.

In contrast with these three cases, the $J_3=2.5$ phase presents energy modes in the gap that do not connect the negative and positive energy bands. By increasing the value of $J_3$ further, one does not find any additional closings in the energy spectrum, and furthermore $\nu_{3D}=0$ and $\nu_{2D}=1$ (for $J_1=K_z=0$). However, by using Eq. \ref{nu1d}, we find that $\nu_{1D}=-1$ when we set $J_3=K_z=J_0=0$. Hence, the system is essentially in a state of weak topological phase of coupled topological  wires. The fact that there is no spectral flow in the energy is because, in the limit of weak coupling, the boundary states of the 1D wires do not have spectral flow anyway. Thus, when coupled the boundary states will not  generically disperse sufficiently strong to reach the bulk energy bands, and even if they did they would not spectrally connect the lower band to the upper band. This behavior of course has its counterpart in the entanglement spectrum where the entanglement modes cross $1/2$ but do not flow between $0$ and $1$.

As we mentioned earlier, the abelian and non-abelian phases of the Kitaev model can be distinguished in their entanglement properties by additional  contributions that appear in the non-abelian phases when the system has edge states. In the present case of the RKD model, if we were to compute the entanglement using open boundary conditions in the $\mathbf{a}_3$ direction, the degenerate zero modes from the two surfaces will contribute additional entanglement to the system, similar to what happens in the 2D Kitaev model. However, whereas in the Kitaev model such additional contribution to the entanglement was linked to the types of excitations in the system by Yao and Qi, in the RKD model the connection to excitations is not clear. We leave this question for future work.

\begin{figure}
\begin{center}
\includegraphics[trim =2cm 1cm 0cm 0cm,scale=0.30]{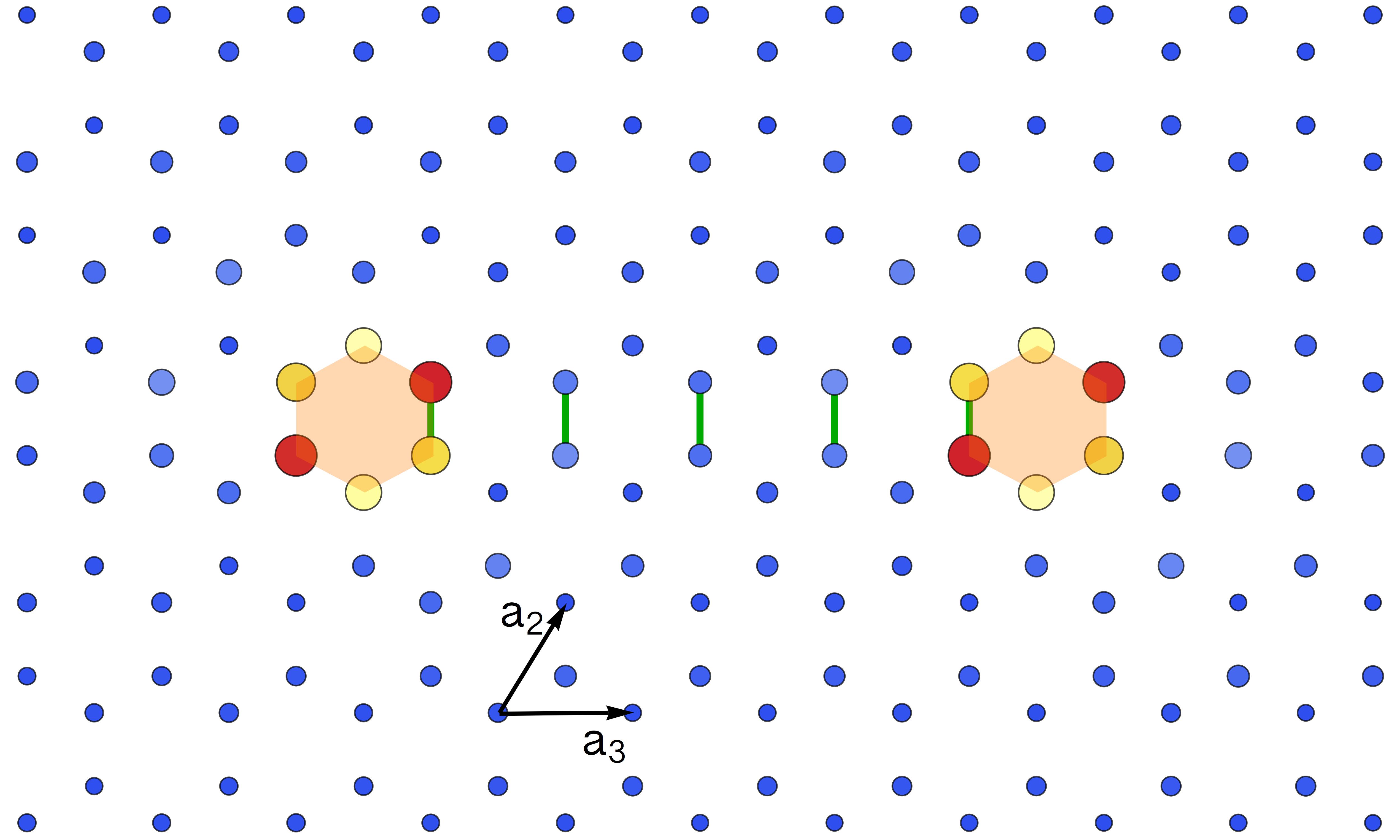}
\caption{ Density profile of the zero modes at $k_1=0$ when $J=1.5$ in one of the layers perpendicular to the $\mathbf{s}_1$ direction. The magnitude of the density is represented by the color and size of the circles at each lattice, with the warmer colors and larger size denoting higher density. The green lines are the links for which the sign is flipped with respect to the ground state configuration of the $Z_2$ gauge field. The shaded hexagons show where the vortices are realized. The vortex lines extend into the plane along the $\mathbf{a}_1$ direction. The $\mathbf{a}_2$ and $\mathbf{a}_3$ directions are shown by the black vectors. }\label{conf}
\end{center}
\end{figure}

\begin{figure}
\begin{center}
\includegraphics[trim =2cm 1cm 0cm 0cm,scale=0.302]{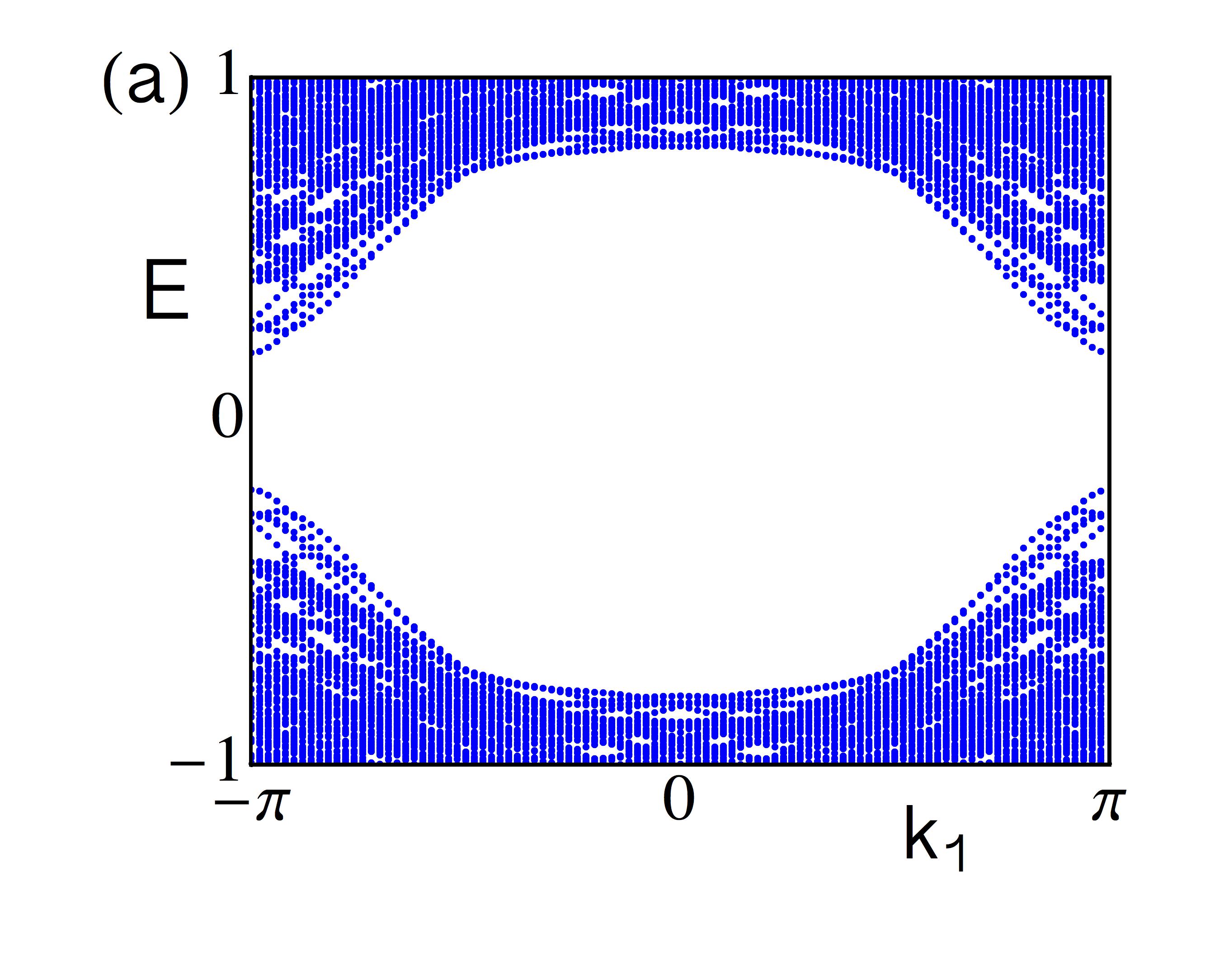}
\includegraphics[trim =2cm 1cm 0cm 0cm,scale=0.302]{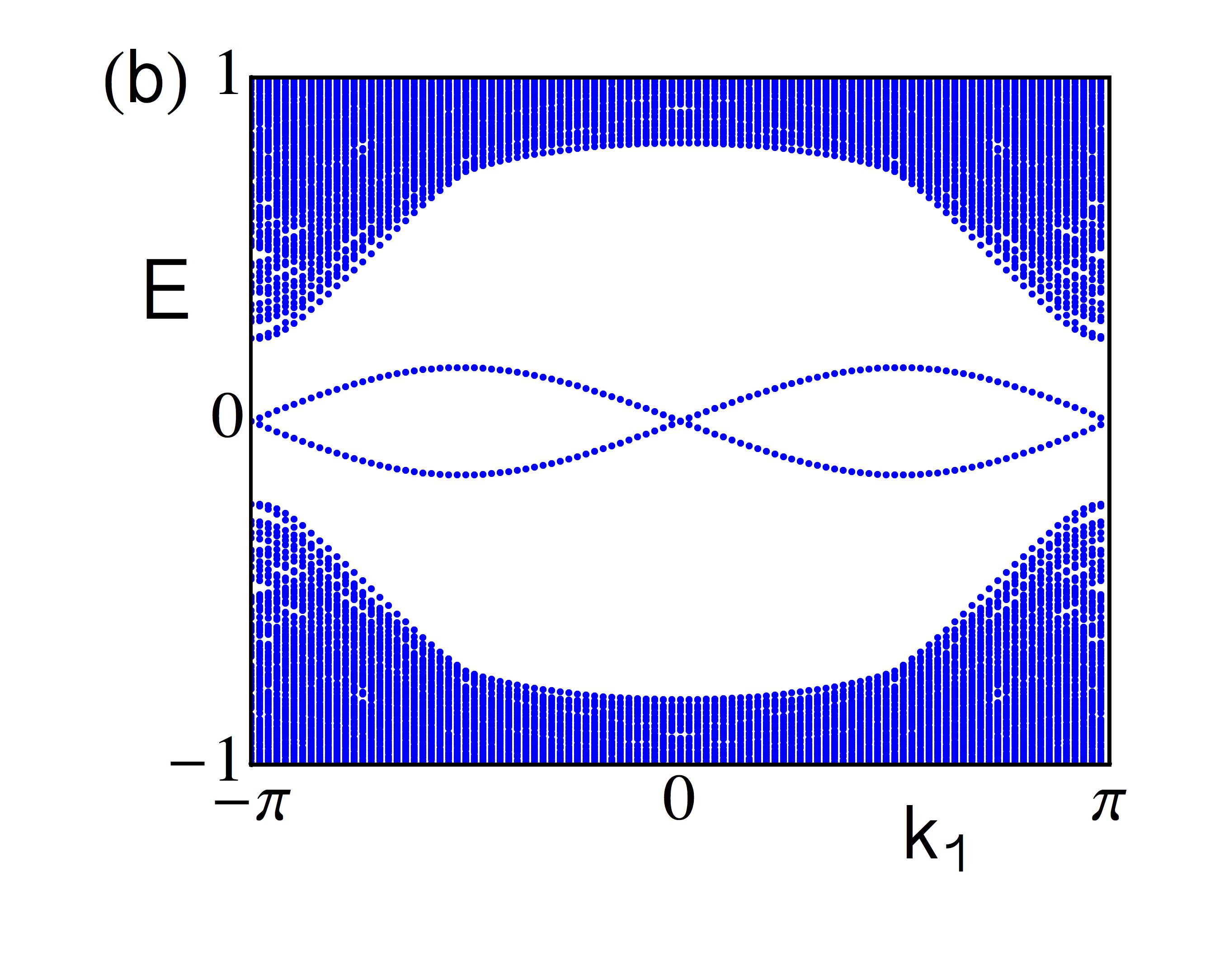}
\includegraphics[trim =2cm 2cm 0cm 0cm,scale=0.302]{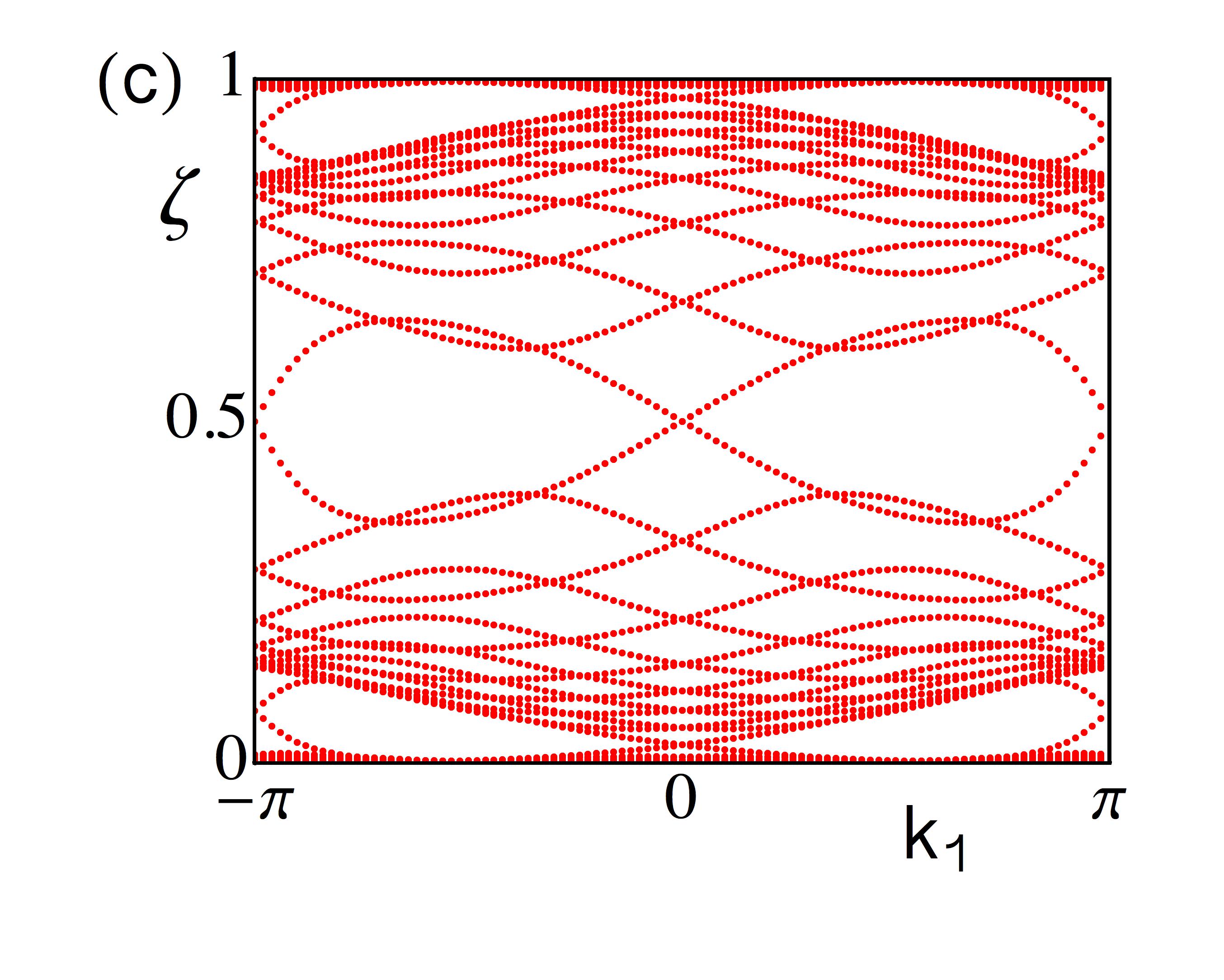}
\includegraphics[trim =2cm 2cm 0cm 0cm,scale=0.302]{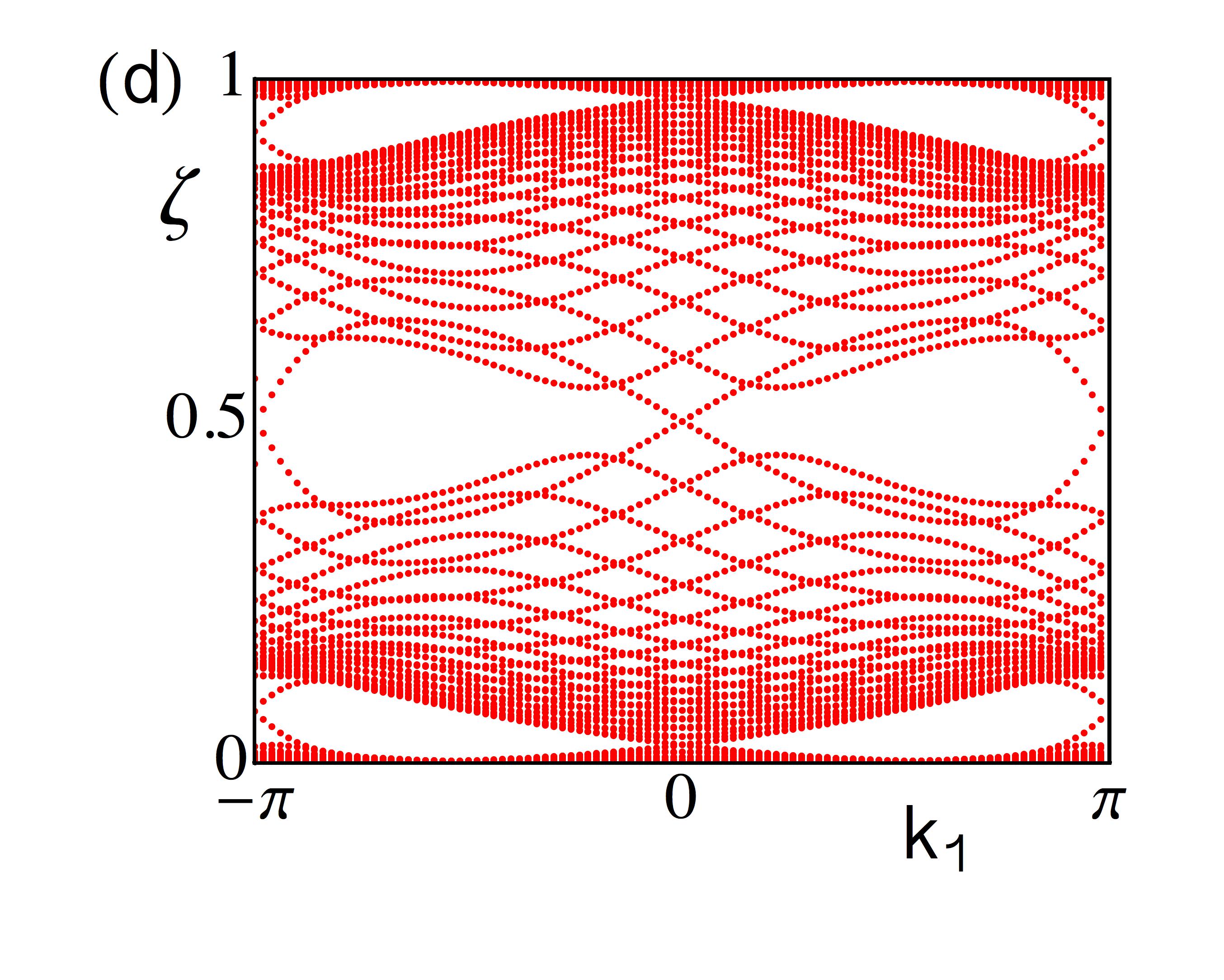}

\vspace{0.5cm}
\includegraphics[trim =2cm 2cm 0cm 0cm,scale=0.35]{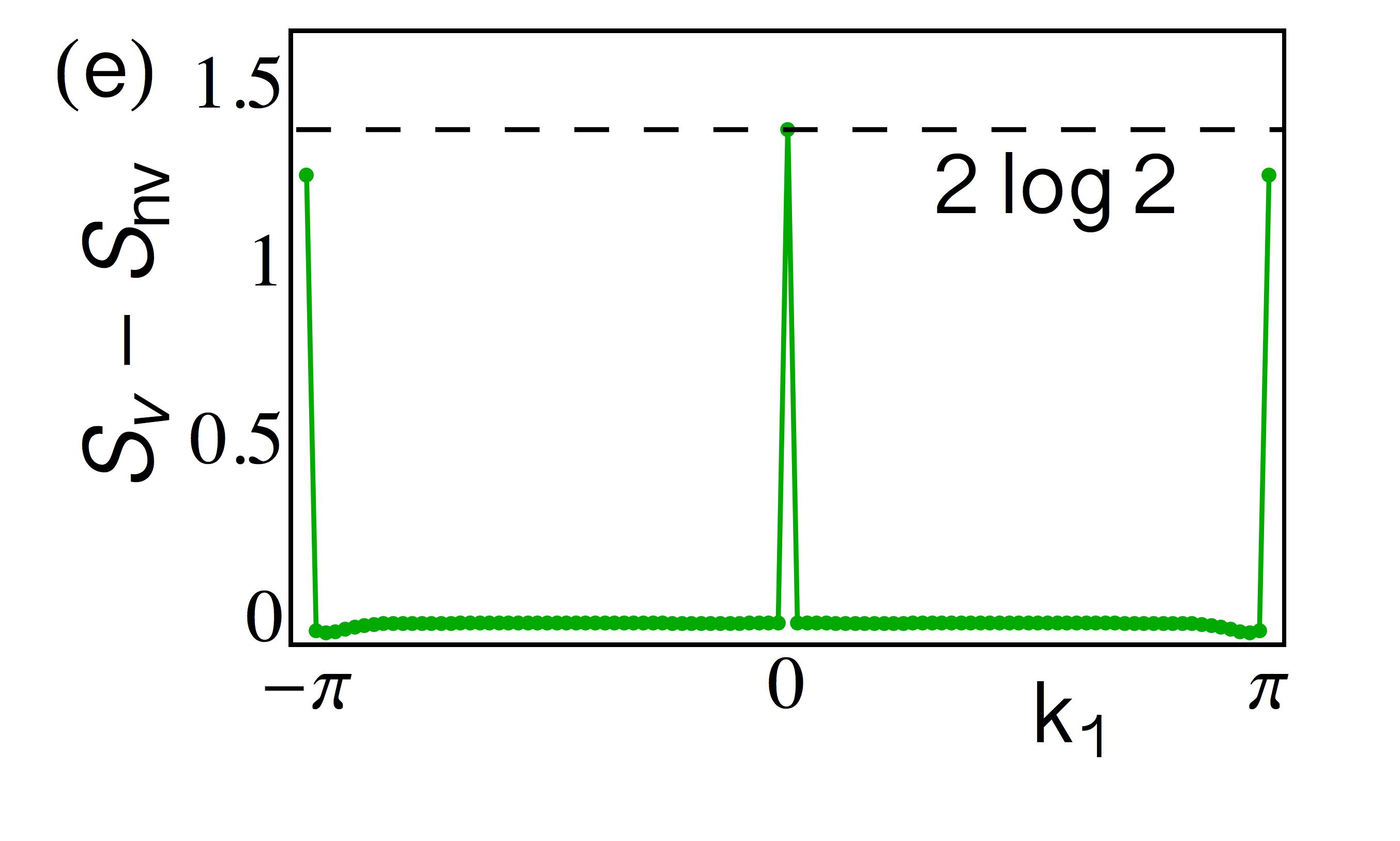}
\caption{ Energy (a,b) and entanglement (c,d) spectra for $J_3=1.5$ strong topological phase before and after adding the vortex lines, respectively. Figure (e) shows the difference in entanglement entropy between both cases, illustrating the additional entanglement obtained by the crossings of the Majorana modes trapped in the vortex lines.}\label{vortexweak}
\end{center}
\end{figure}

\subsection{Entanglement  arising from vortices in the $Z_2$ gauge field}

We now discuss the case of the entanglement that arises from introducing vortex configurations in the $Z_2$ gauge field. We have found that the derivation of the factorization of the density matrix continues to hold, regardless of whether the $Z_2$ has vortex configurations.  Furthermore, the $Z_2$ gauge field will continue to contribute the same amount of entanglement entropy as it did for the ground state. Consequently, any change to the entanglement of the system will arise from the Majorana modes that are trapped by the $Z_2$ vortices. This allows us to easily study the entanglement in the presence of the $Z_2$ flux excitations of the gauge field.

To generate a vortex, one changes the signs of the links in such a way that the product of the links around the hexagon leads to $\prod_{\bar{ij}\in \text{hex}}u_{ij}=-1$. To simplify the discussion, we will consider periodic boundary conditions in all directions and introduce two vortex lines parallel to the $\mathbf{a}_1$ direction.

\begin{figure}
\begin{center}
\includegraphics[trim =2cm 1cm 0cm 0cm,scale=0.302]{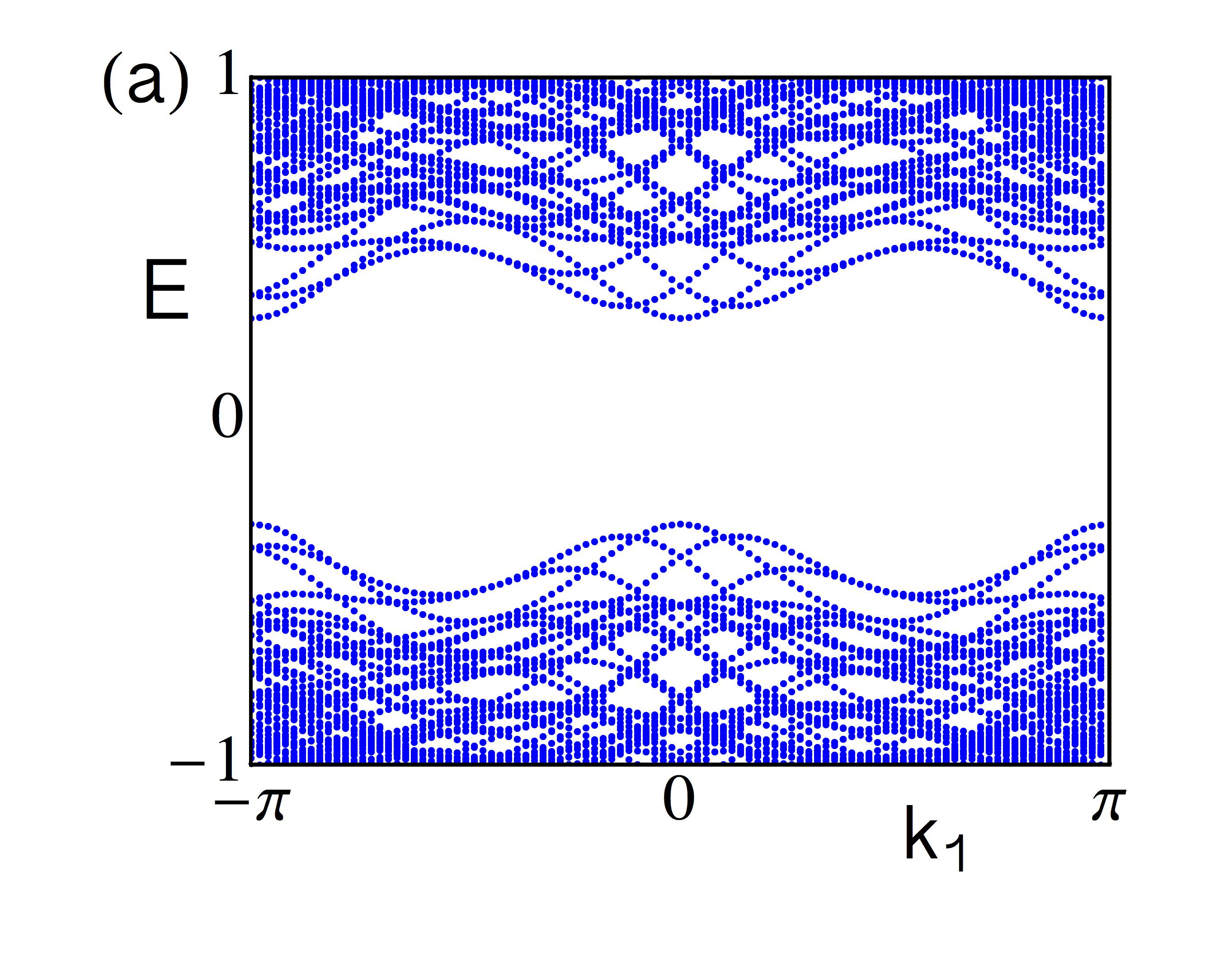}
\includegraphics[trim =2cm 1cm 0cm 0cm,scale=0.302]{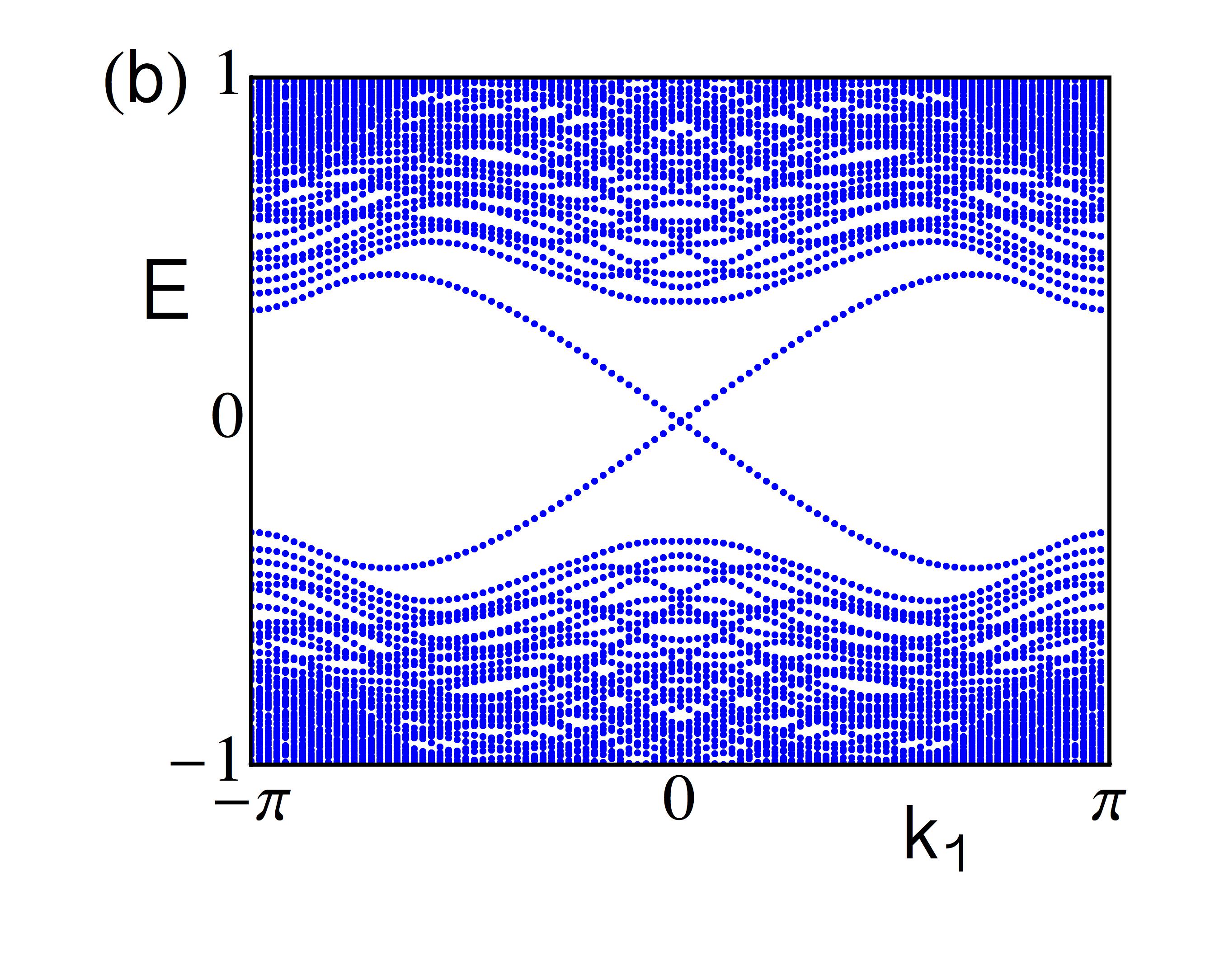}
\includegraphics[trim =2cm 2cm 0cm 0cm,scale=0.302]{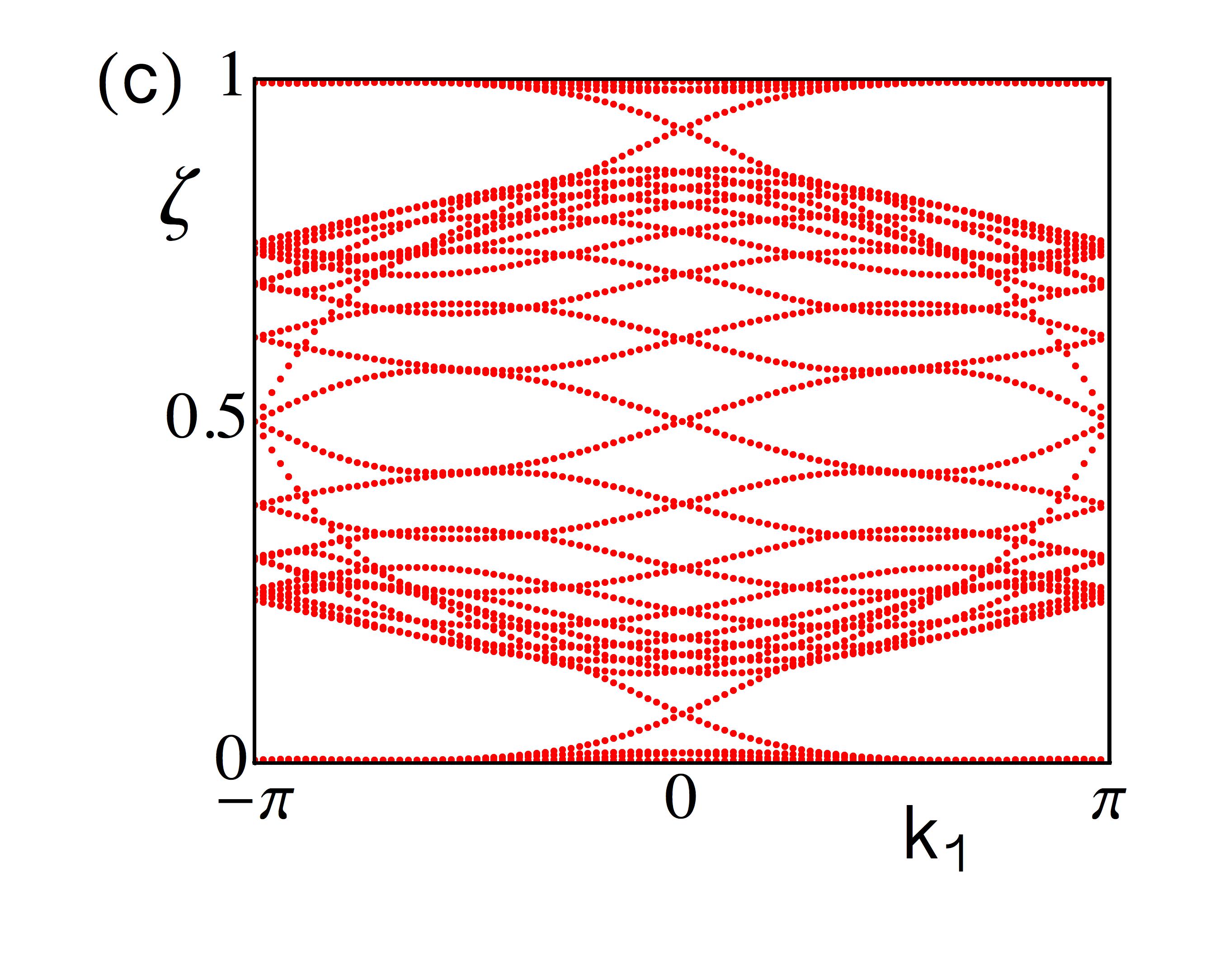}
\includegraphics[trim =2cm 2cm 0cm 0cm,scale=0.302]{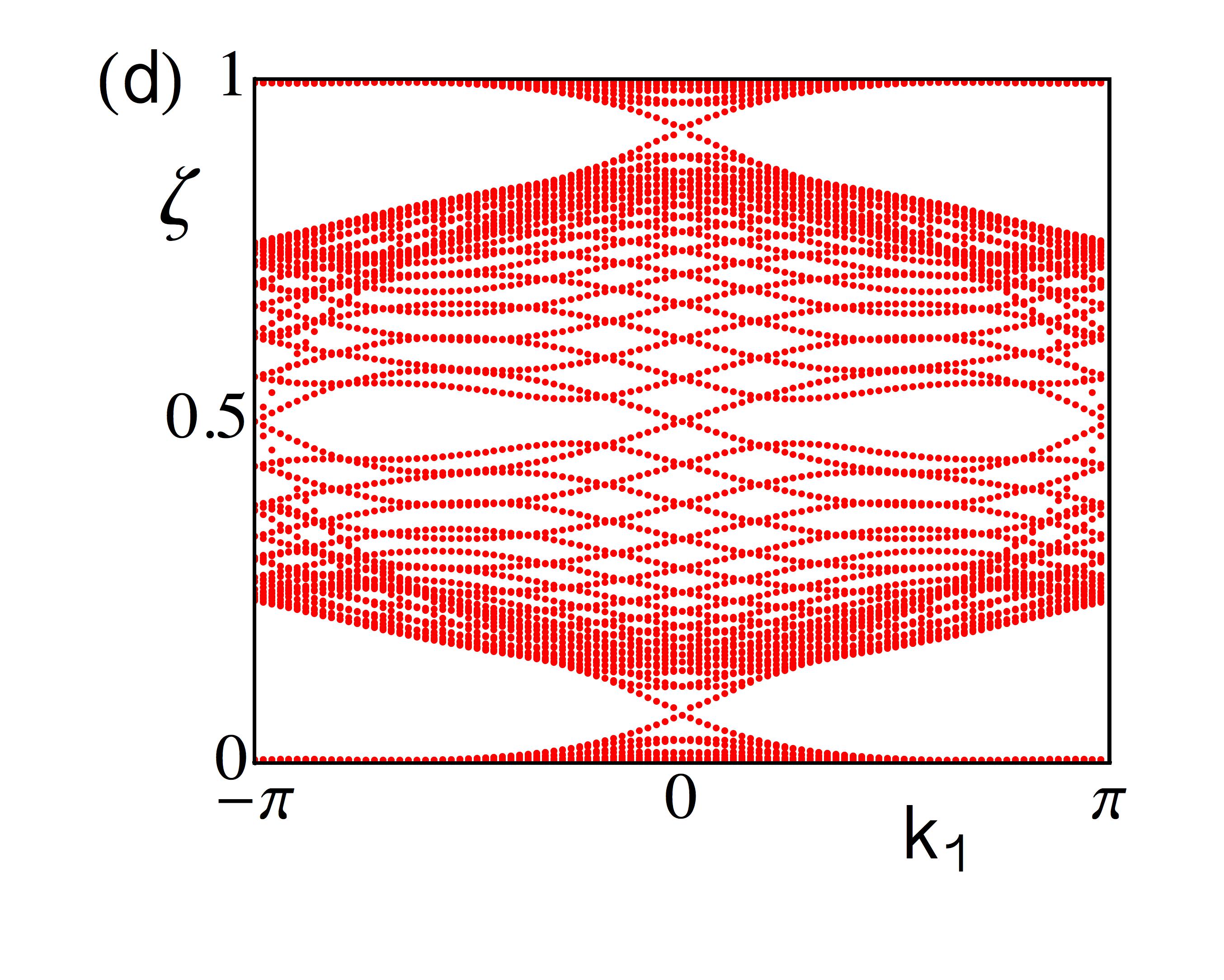}

\vspace{0.5cm}
\includegraphics[trim =2cm 2cm 0cm 0cm,scale=0.35]{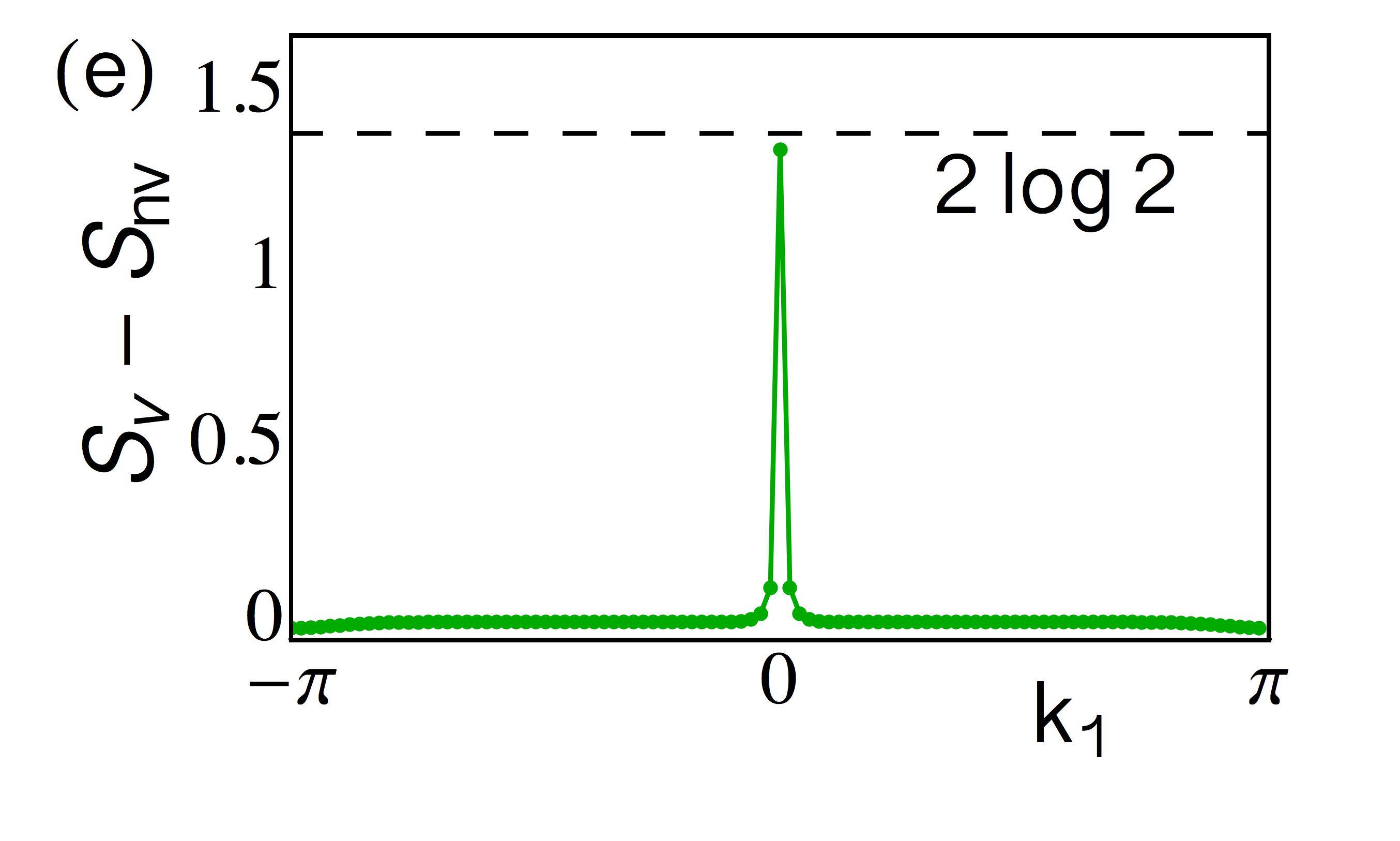}
\caption{ Energy (a,b) and entanglement (c,d) spectra for $J_3=1.0$ weak topological phase before and after adding the vortex lines, respectively. Figure (e) shows the difference in entanglement entropy between both cases, illustrating the additional entanglement obtained by the single crossing of the Majorana modes trapped in the vortex lines.}\label{vortexstrong}
\end{center}
\end{figure}

Consider a system of dimensions $N_{1,2,3}$ along the $\mathbf{a}_{1,2,3}$ directions respectively. The configuration we choose here of the $Z_2$ gauge field corresponds to assigning a minus sign to the link operators of the form $\hat{u}_{r_A, r_A-s_0}$, such that $r_A=n_1 \mathbf{a}_1+(N_2/2)\mathbf{a}_2+n_3 \mathbf{a}_3$, with $n_1=1, \ldots, N_1$ and $n_3=N_3/4,\ldots, 3N_3/4$. To illustrate this $Z_2$ configuration, we show the pattern of signs in Fig. \ref{conf}. for one of the layers with normal vector $[1\,1\,1].$  Because of this choice of $Z_2$ gauge field, the vortex lines are threaded through the shaded hexagons and extend along the $\mathbf{a}_1$ direction.  Once this choice of link values is set, both the nearest and next-nearest neighbor tunneling terms have to be changed accordingly because both types of tunneling are written in terms of link operators.

The Majorana fermions will feel the presence of the vortex lines through the phases of the hopping parameters. Such vortex lines can induce states in the gap of the system when the bulk is topologically nontrivial. Even though these states will be localized to either region $A$ or $B$, they can contribute to the entanglement of the system.  For simplicity, we consider two of the cases of the previous section, namely the strong topological state at $J_3=1.5$ and the weak topological state at $J_3=1.0$. As we now discuss, there is a clear distinction between the entanglement modes of both cases when vortices are introduced. 

 In Figs. \ref{vortexweak}a,b we show the energy spectrum with and without the vortex lines when $J_3=1.0$. The presence of the vortex lines induces  doubly degenerate Majorana branches that cross at $k_1=0$ and  $k_1=\pi$. The corresponding entanglement spectrum is shown in Figs. \ref{vortexweak}c,d. Similarly, in Figs. \ref{vortexstrong}a,b we show the energy spectrum with and without the vortex lines when $J_3=1.5$. In this case, the doubly degenerate Majorana branches cross at the single point $k_1=0$. The corresponding entanglement spectrum is shown in Figs. \ref{vortexstrong}c,d. The double degeneracy is due to the presence of two vortex lines.

The behavior we observe here can be understood from the arguments presented in \cite{Teo2010}. Vortex lines can be seen as one-dimensional defects in three-dimensional systems that belong, in this case, to class DIII. It was shown in \cite{Teo2010} that under these circumstances, there is a $Z_2$ classification of the state. The invariant associated with this classification determines the stability of gapless Majorana modes that propagate along the vortex line.  

We can infer from the number of crossings in the energy and entanglement spectrum that the types of gapless Majorana modes we have obtained have a different $Z_2$ invariant for the $J_3=1.0$ and $J_3=1.5$ cases. The weak topological state presents gapless Majorana modes that cross zero energy an even number of times, whereas in the strong topological state the Majorana mode crosses zero an odd number of times. This feature is also present in the entanglement spectrum. Such crossings lead to additional entanglement in the system, with respect to the case of no vortices and with periodic boundary conditions.  We further emphasize this point in Fig.\ref{vortexweak}e and Fig.\ref{vortexstrong}e by showing the difference $S_{v}(k_1)-S_{nv}(k_1)$, where $S_{v}(S_{nv})$ is the entanglement entropy of region $A$ when the vortex lines are present (absent). There is an additional contribution that is approximately $2\log 2$  for each of the crossings obtained in the entanglement spectrum. In the weak topological state this contribution comes from the two crossings $k_1=0,\pi$, whereas in the strong topological state this occurs only at $k_1=0$.

\section{Conclusions}

In this work, we have explored the entanglement properties of a three dimensional generalization of the Kitaev model proposed by Ryu. We have shown that the entanglement entropy separates into a contribution from the $Z_2$ gauge field and the Majorana degrees of freedom, in the same way that it occurs for the Kitaev model. We took advantage of these properties to explore the behavior of the entanglement spectrum of both weak and strong topological phases of the model proposed by Ryu. Finally, we considered the effect of introducing vortex lines in the $Z_2$ gauge field, which lead to additional contributions to the entanglement entropy arising from gapless Majorana modes trapped in the vortices.

\section{Acknowledgements}

This work was supported by ONR award N0014-12-1-0935. We acknowledge a useful conversation with S. Ryu and the support of the UIUC ICMT.

\appendix

\section{$Z_2$ topological invariant for 2D and 1D systems in class DIII}\label{2dnu}

Here we obtain the topological invariant of the layers perpendicular to the $\mathbf{s}_1$ direction when $K_z=0$ and $J_1=0$. We follow the line of reasoning in \cite{Ryu2012}. In this limit, the 3D model is effectively a set of two-dimensional systems in class DIII, each of which is classified by a $Z_2$ invariant. We can thus use the Fu-Kane formula
\begin{equation}
\nu_{2D}=\prod_{\mathbf{k} \in  \text{TRIM}}\frac{\sqrt{ \det \left(w(\mathbf{k})\right)}}{\text{Pf}\left[w(\mathbf{k})\right]},
\end{equation}
where $w_{nm}(\mathbf{k})=\bra{u_{n}(-\mathbf{k})}T\ket{u_{m}(\mathbf{k})}$ and $\text{Pf}[w]$ is the Pfaffian of the matrix $w$.

The calculation of this invariant can be simplified by exploiting the chiral symmetry of the model, with the chiral operator given by $S=s^{y} c^{z}$.  In the basis in which $S$ is diagonal, the operator $Q(\mathbf{k})=2P(\mathbf{k})-1$ can be written in block off-diagonal form (where $P(\mathbf{k})$ is the projection operator into the occupied states). Let us then define the block in the block-off diagonal as $q(\mathbf{k})$. Then through the unitary transformation that diagonalizes $S$, namely
\begin{equation}
U=\frac{1}{\sqrt{2}}\left(
\begin{array}{cccc}
 0 & i & 0 & -i \\
 0 & 1 & 0 & 1 \\
 -i & 0 & i & 0 \\
 1 & 0 & 1 & 0 \\
\end{array}
\right),
\end{equation}
 the $q(\mathbf{k})$ matrix is given by
\begin{equation}
q(\mathbf{k})=\frac{1}{\epsilon(\mathbf{k})}\left(
\begin{array}{cc}
 i \Theta_x+\Theta_z & -\text{Im}\Phi -i\text{Re}\Phi\\
-\text{Im}\Phi +i\text{Re}\Phi & -i \Theta_x-\Theta_z
\end{array}
\right),
\end{equation}
where $\epsilon(\mathbf{k})=\sqrt{\vert \Phi\vert^2+\Theta^{x 2}+\Theta^{z 2}}$, with $\vert \Phi\vert^2=(\text{Re}\Phi)^2+(\text{Im}\Phi)^2$. In this form, we can can write the single-particle eigenstates as $\ket{u(\mathbf{k})}=\frac{1}{\sqrt{2}}\left(n_a\, q^{\dagger}(\mathbf{k}) n_a\right)^T$, where $n_{1}=(1,\,0)^T$ and $n_{2}=(0,\,1)^T$. From this we then obtain 
\begin{equation}
w_{ab}=i q^{\dagger}_{ab}(\mathbf{k}).
\end{equation}
The TRIM are given by $\mathbf{k}=\frac{1}{2}\left(m_2 b_{2}+m_3 b_{3}\right)$, where $\mathbf{b}_i \cdot \mathbf{a}_j=2\pi \delta _{ij}$ and $(m_2,m_3)=(0,0), (0,1),(1,0),(1,1)$. By setting $K_z=J_1=0$ (the limit of decoupled layers),  we get $w_{11}=w_{22}=0$ and $w_{12}=-w_{21}=J_2+J_0 \cos \pi m_2+J_3 \cos \pi m_3$, which implies  
\begin{equation}
\nu_{2D}=\text{sign}(J_0+J_2+J_3) \text{sign}(J_0-J_2+J_3) \text{sign}(-J_0+J_2+J_3) \text{sign}(J_0+J_2-J_3).
\end{equation}

The corresponding expression for the $1D$ invariant  can be computed in the same way, except that now there are only two time-reversal invariant momenta to consider and we set $J_0=0$. In this case $w_{11}=w_{22}=0$ and $w_{12}=-w_{21}=J_2+J_3 \cos \pi m_3$ with $m_3=0,\pi$. This then leads to 
\begin{equation}
\nu_{1D}=\text{sign}(J_2+J_3)\text{sign}(J_2-J_3). \nonumber
\end{equation}

\section{ Derivation of the factorization of the density matrix in the RKD model}\label{proof}

In this appendix, we will derive the factorization of $\text{Tr}_A\left[\rho_A^n\right]$ in terms of gauge field and Majorana parts, where $\rho_A$ is the reduced density matrix of region $A$. In order to do this, we will first obtain a simplified expression for $\rho_A=\text{Tr}_B\left[\ket{\psi}\bra{\psi}\right]$. We start by writing the explicit eigenstate of the system as
\begin{equation}
\ket{\psi}= \sqrt{\frac{1}{2^{1-N}}}\prod_{j}\left(\frac{1+D_j}{2}\right)\ket{u}\otimes \ket{\phi(u)}, \label{proj}
\end{equation}
where $N$ is the number of sites in the lattice. This state represents a possible eigenstate for a given configuration of the flux through each hexagon of the diamond lattice.

To simplify the trace over the $Z_2$ gauge field, we rewrite the links that cross the entanglement cut in terms of new states of links that exist exclusively in either region $A$ or $B$. Let us assume that there are $2L$ links that cut through the entanglement cut (the odd case can also be considered with some modifications).  We write the gauge field configuration in the factorized form
\begin{equation}
\ket{u}=\ket{u_A}\ket{u_p}\ket{u_B}, \nonumber
\end{equation}
where $\ket{u_A}$ and $\ket{u_B}$ describe all the links in regions $A$ and $B$, respectively, and the state $\ket{u_p}$ describes the links that cross the entanglement cut. We denote the link operators that cross the entanglement cut by $u_{a_n b_n}$, where $a(b)$ label region  $A(B)$ and  $n,m =1\ldots 2L$. We seek a basis transformation between the set of link operators   $\{u_{a_n b_n}=i\lambda_{a_n}^\alpha \lambda_{b_n}^\alpha, u_{a_m b_m}=i\lambda_{a_m}^\beta \lambda_{b_m}^\beta\}$ and $\{w_{a_n a_m}=i\lambda_{a_n}^\alpha \lambda_{a_m}^\beta,w_{b_m b_n}=i\lambda_{b_m}^\beta\lambda_{b_n}^\alpha\}$. This new set of link operators does not pierce through the entanglement cut.

One finds that this basis transformation leads to 
\begin{equation}
\ket{u_p}=\frac{1}{\sqrt{2^L}}\sum_{w_A, w_B} c_{w_{AB}}\ket{w_A, w_B},\label{wab1}
\end{equation}
where $\ket{u_p}$ and $\ket{w_A, w_B}$ denote collectively all the $2L$ links that are involved in the basis transformation. The sum runs over all possible configurations of the $w_A$ links. For each of these configurations, the basis transformation dictates a corresponding configuration of the $w_B$ links. The coefficients of the expansion take the values $c_{w_{AB}}=\pm1$, depending on the particular configuration of the original $Z_2$ gauge field.

We can now write the physical state as 
\begin{eqnarray}
\ket{\psi}=\frac{1}{\sqrt{2^{N+L+1}}}\sum_{w_{AB},g}\prod_{i \in g} D_i c_{w_{AB}} \ket{u_A w_A; u_B, w_B}\ket{\phi(u)}. \nonumber
\end{eqnarray}
The product $\prod_{i \in g} D_i$ can be factorized as
\begin{eqnarray}
\prod_{i \in g} D_i=\prod_{i \in g}\left(i \lambda^0_i \lambda^1_i\lambda^2_i\lambda^3_i\lambda^4_i\lambda^5_i\right)=X_{g_A}X_{g_B}Y_{g_A}Y_{g_B}\nonumber
\end{eqnarray}
Here, we defined the operators $X_g=\prod_{i\in g} \lambda^0_i \lambda^1_i \lambda^2_i \lambda^3_i$ and $Y_g=\prod_{i\in g} i \lambda^4_i \lambda^5_i$. These operators can now be distributed to act on the $Z_2$ field and Majorana parts respectively
\begin{eqnarray}
\ket{\psi}=\frac{1}{\sqrt{2^{N+L+1}}}\sum_{g, w_{AB}} c_{w_{AB}} &&X_{g_B}\ket{u_B w_B}X_{g_A}\ket{u_A w_A} Y_{g_A}Y_{g_B}\ket{\phi(u)},\nonumber
\end{eqnarray}
which leads to the following form of the reduced density matrix of region $A$:
\begin{eqnarray}
\fl \rho_A=\frac{1}{2^{N+L+1}} \sum_{\substack{g, w_{A}\\g', w'_{A}}}c_{w_{AB}} c^*_{w'_{AB}}\left( X_{g_A}\ket{u_A w_{A}}\bra{u_A w'_{A}}X^\dagger_{g'_A}\right)\nonumber\\
\times\text{Tr}_B\left[\left( X_{g_B}\ket{u_B w_{B}}\bra{u_B w'_{B}}X^\dagger_{g'_B}\right)\left( Y_{g_A}Y_{g_B}\ket{\phi(u)}\bra{\phi(u)}Y^\dagger_{g'_B}Y^\dagger_{g'_A}\right)\right]\nonumber\\
\fl= \frac{1}{2^{N+L+1}} \sum_{\substack{g, w_{A}\\g', w'_{A} }}  c_{w_{AB}} c^*_{w'_{AB}}\left( X_{g_A}\ket{u_A w_{A}}\bra{u_A w'_{A}}X_{g'_A}\right)\text{Tr}_{B,G}\left[X_{g_B}\ket{u_B w_{B}}\bra{u_B w'_{B}}X_{g'_B}\right]\nonumber\\
\times Y_{g_A}\text{Tr}_{B,F}\left[ Y_{g_B}\ket{\phi(u)}\bra{\phi(u)}Y_{g'_B}\right]Y_{g'_A}\nonumber\\
\fl= \frac{1}{2^{N+L+1}}\sum_{\substack{g, w_{A}\\g', w'_{A}}}  c_{w_{AB}} c^*_{w'_{AB}}\left( X_{g_A}\ket{u_A w_{A}}\bra{u_A w'_{A}}X_{g'_A}\right)\left(\bra{u_B w'_{B}}X_{g'_B}X_{g_B}\ket{u_B w_{B}}\right)\nonumber\\
\times Y_{g_A}\text{Tr}_{B,F}\left[\left( Y_{g_B}\ket{\phi(u)}\bra{\phi(u)}Y_{g'_B}\right)\right]Y_{g'_A}\label{rhoa_1}, \nonumber
\end{eqnarray}
where the subindices $G$ and $F$ stand for tracing over gauge and fermion degrees of freedom.

 In order to further simplify the expression for the reduced density matrix we must calculate the matrix element $\bra{u_B w'_{B}}X_{g'_B}X_{g_B}\ket{u_B w_{B}}$. Note that the $\lambda^p_i$ operators that compose $X_g$ will switch the signs of whatever configuration $\ket{u_B w_B}$ has for each of the four links associated to the sites in the set $g'_B \cup g_B$. Hence, in general the re-configured ket $X_{g_B}\ket{u_B w_{B}}$ will not match the bra $\bra{u_B w'_{B}}X_{g'_B}$, unless either the set $g'_B \cup g_B$ covers all the sites in $B$ i.e. $g'_B \cup g_B=B$, or unless both products cover exactly the same sites i.e. $g'_B=g_B$. These two possibilities then lead to 
\begin{eqnarray}
\bra{u_B w'_{B}}X_{g'_B}X_{g_B}\ket{u_B w_{B}}&=& \delta_{w_B w'_B} \left(\delta_{g'_B g_B}+\bra{u_B w_{B}}X_{B}\ket{u_B w_{B}} \delta_{g'_B+g_B,\,B}\right)\nonumber\\
&=&\delta_{w_B w'_B} \left(\delta_{g'_B g_B}+\prod_{\bar{ij}}u_{ij }\prod_{n}w_{B_n} \delta_{g'_B+g_B,\,B}\right)\nonumber\\
&=&\delta_{w_B w'_B} \left(\delta_{g'_B g_B}+x_B(w_B) \delta_{g'_B+g_B,\,B}\right),
\end{eqnarray}
where we used the fact that $X_B$ can be rearranged as a product of link operators in $B$ and we defined $x_B(w_B)=\prod_{\overline{ij}}u_{ij}\prod_{n}w_{B_n}$. Although there can be an overall minus sign in $x_B(w_B)$ due to the rearrangement done to form the link operators (depending on the number of sites), as we will see below this sign would not affect the end result. We thus obtain the simplified expression
\begin{eqnarray}
\fl\rho_A = \sum_{\substack{g_A g_B w_A\\ g'_A g'_B w'_A }}\frac{c_{w_{AB}} c^*_{w'_{AB}}}{2^{N+L+1}}X_{g_{A}}\ket{u_A w_A}\bra{u_A w'_A}X_{g'_A} \delta_{w_B w'_B} \left(\delta_{g'_B g_B}+x_B(w_B)\delta_{g'_B+g_B,\,B}\right)Y_{g_A}\nonumber\\ 
\times\text{Tr}_{B,F}\left[ Y_{g_B}\ket{\phi(u)}\bra{\phi(u)}Y_{g'_B}\right]Y_{g'_A} \nonumber\\
\fl= \sum_{\substack{g_A g_B w_A \\ g'_A g'_B}}\frac{\vert c_{w_{AB}}\vert^2}{2^{N+L+1}}X_{g_{A}}\ket{u_A w_A}\bra{u_A w_A}X_{g'_A}  Y_{g_A}\text{Tr}_{B,F}\left[\ket{\phi(u)}\bra{\phi(u)}Y_{g'_B} Y_{g_B}\left(\delta_{g'_B g_B}+x_B(w_B)\delta_{g'_B+g_B,\,B}\right)\right]Y_{g'_A} \nonumber\\
\fl=\frac{1}{2^{N_A+L}}\sum_{\substack{g_A g'_A w_A}} X_{g_A}\ket{u_A w_A}\bra{u_A w'_A}X_{g'_A} Y_{g_A} \text{Tr}_{B,F}\left[\ket{\phi(u)}\bra{\phi(u)}\left(\frac{1+x_B(w_B)\eta_B}{2}\right)\right]Y_{g'_A}\nonumber\\
\fl=\frac{1}{2^{N_A+L}}\sum_{g_A g'_A w_A}X_{g_A}\ket{u_A w_A}\bra{u_A w_A}X_{g'_A} Y_{g_A} \rho_{A,F}^{x_B(w_B)}Y_{g'_A}
\end{eqnarray}
where we defined $\eta_B=Y_{B-g_B}Y_{g_B}=\prod_{i\in B}\left(i\lambda_i^4 \lambda_i^5\right)$ and  $\rho_{A,F}^{x_B(w_B)}=\text{Tr}\left[\ket{\phi(u)}\bra{\phi(u)}\left(\frac{1+x_B(w_B)\eta_B}{2}\right)\right]$. We can calculate the second power of the reduced density matrix
\begin{small}
\begin{eqnarray}
\fl\rho_A^2=\frac{1}{2^{2(N_A+L)}}\sum_{\substack{g_A g'_A w_A\\h_A h'_A v_A}}X_{g_A}\ket{u_A w_A}\bra{u_A w_A}X_{g'_A} Y_{g_A}\rho_{A,F}^{x_B(w_B)}Y_{g'_A}X_{h_A}\ket{u_A v_A}\bra{u_A v_A}X_{h'_A} Y_{h_A}\rho_{A,F}^{x_B(v_B)}Y_{h'_A}\nonumber\\
\fl=\frac{1}{2^{2(N_A+L)}}\sum_{\substack{g_A g'_A w_A\\h_A h'_A v_A}}X_{g_A}\ket{u_A w_A}\bra{u_A v_A}X_{h'_A}\left(\bra{u_A w_A}X_{g'_A}X_{h_A}\ket{u_A v_A}\right) Y_{g_A}\rho_{A,F}^{x_B(w_B)}Y_{g'_A} Y_{h_A}\rho_{A,F}^{x_B(v_B)}Y_{h'_A}\nonumber\\
\fl=\frac{1}{2^{2(N_A+L)}}\sum_{\substack{g_A g'_A w_A\\h_A h'_A v_A}}X_{g_A}\ket{u_A w_A}\bra{u_A v_A}X_{h'_A}\delta_{w_A, v_A}\left(\delta_{g'_A h_A}+ x_A(w_A) \delta_{g'_A +h_A, A}\right) Y_{g_A}\rho_{A,F}^{x_B(w_B)}Y_{g'_A} Y_{h_A}\rho_{A,F}^{x_B(v_B)}Y_{h'_A}\nonumber\\
\fl=\frac{1}{2^{2(N_A+L)}}\sum_{\substack{g_A w_A\\h_A h'_A }}X_{g_A}\ket{u_A w_A}\bra{u_A w_A}X_{h'_A}Y_{g_A}\rho_{A,F}^{x_B(w_B)}\left(1+x_A(w_A) Y_{A-h_A} Y_{h_A}\right)\rho_{A,F}^{x_B(w_B)}Y_{h'_A}\nonumber\\
\fl=\frac{1}{2^{2(N_A+L)-1}}\sum_{\substack{g_A  w_A h'_A }}X_{g_A}\ket{u_A w_A}\bra{u_A w_A}X_{h'_A}Y_{g_A}\rho_{A,F}^{x_B(w_B)}\left(\frac{1+ x_A(w_A) \eta_A}{2}\right)\rho_{A,F}^{x_B(w_B)}Y_{h'_A}\nonumber\\
\fl=\frac{1}{2^{N_A+2L-1}}\sum_{g_A  w_A h'_A  }X_{g_A}\ket{u_A w_A}\bra{u_A w_A}X_{h'_A}Y_{g_A}\rho_{A,F}^{x_B(w_B)}P_{A,F}^{x_A(w_A)}\rho_{A,F}^{x_B(w_B)}Y_{h'_A}\nonumber
\end{eqnarray}
\end{small}
where we defined $P_{A,F}^{x_A(w_A)}=\left(\frac{1+ x_A(w_A) \eta_A}{2}\right)$. By computing $\rho_A^3$ and successive powers, the same matrix elements that we have used so far appear iteratively, and through the same type of simplification we have employed one can thus infer the $n$-th power of the reduced density matrix to be:
\begin{eqnarray}
\fl\rho_A^n=&&\frac{1}{2^{N_A+nL-(n-1)}}\sum_{g_A w_A h'_A }X_{g_A}\ket{u_A w_A}\bra{u_A w_A}X_{h'_A} Y_{g_A}\rho_{A,F}^{x_B(w_B)}\left(P_{A,F}^{x_A(w_A)}\rho_{A,F}^{x_B(w_B)}\right)^{n-1} Y_{h'_A}.\nonumber
\end{eqnarray}
With this general form in hand, we can finally obtain an expression for the trace of $\rho_A^n$ in region $A$:
\begin{eqnarray}
\text{Tr}_A\left[\rho_A^n\right]&=&\frac{1}{2^{nL-n}}\sum_{w_A}\text{Tr}_{A,F}\left[ \left(P_{A,F}^{x_A(w_A)}\rho_{A,F}^{x_B(w_B)}\right)^{n}\right]. \nonumber
\end{eqnarray}
We can next divide the sum over the $2^{L-1}$ sets of $w_A$'s that yield $\prod_{n=1}^Lw_{A_n}=+1$ and the $2^{L-1}$ sets that yield $\prod_{n=1}^Lw_{A_n}=-1$. For each of these two cases, we get $\prod_{n=1}^Lw_{B_n}=\sigma$ and $\prod_{n=1}^Lw_{B_n}=-\sigma$ respectively, with $\sigma =\pm 1$ depending on the particular configuration $\ket{u}$ chosen initially for the $Z_2$ gauge field. By defining $p_{A(B)}=\prod_{\bar{ij}\in A(B)}u_{ij}$, we then obtain
\begin{eqnarray}
\text{Tr}_A\left[\rho_A^n\right]&=&\frac{1}{2^{(n-1)(L-1)}}\text{Tr}_{A,F}\left[\left(P_{A,F}^{p_A}\rho_{A,F}^{\sigma p_B}\right)^{n}+\left(P_{A,F}^{-p_A}\rho_{A,F}^{- \sigma p_B}\right)^{n}\right].\nonumber
\end{eqnarray}
Here, we use the fact that 
\begin{eqnarray}
D \ket{u}\otimes \ket{\phi(u)}&=& \prod_{ij} u_{ij}\ket{u}\otimes \eta_A \eta_B \ket{\phi(u)}=\ket{u}\otimes \ket{\phi(u)},\nonumber
\end{eqnarray}
so that $\eta_A \eta_B\ket{\phi(u)}=\prod_{ij} u_{ij} \ket{\phi(u)}$. This means the fermion parity of the state $\ket{\phi(u)}$ is $\prod_{ij} u_{ij}$. Furthermore, note that since $u_{a_n b_n}u_{a_m b_m}=w_{a_n a_m}w_{b_m b_n}$, then $\prod_{ij} u_{ij} =\prod_{ij \in A} u_{ij}  \prod_{n}w_{A_n} \prod_{n}w_{B_n}\prod_{ij \in B} u_{ij}=\sigma p_A p_B$. This implies that the fermion parity of $\ket{\phi(u)}$ is $\sigma p_A p_B$. It follows from this that
\begin{eqnarray}
\rho_{A,F}^{\sigma p_B}P_{A,F}^{-p_A}&=&\text{Tr}_{B,F}\left[\ket{\phi(u)}\bra{\phi(u)} P^{\sigma p_B}_{B,F}\right] P^{-p_A}_{B,F}=0. \nonumber
\end{eqnarray}
because $P^{\sigma p_B}_{B,F}P^{-p_A}_{B,F}$ projects $\ket{\phi(u)}$ into a state of fermion parity $-\sigma p_A p_B$, which is of opposite sign to the parity we found for $\ket{\phi(u)}$.  It thus follows that we can write
\begin{eqnarray}
\fl\text{Tr}_A\left[\rho^n_A\right]=\frac{1}{2^{(n-1)(L-1)}} \text{Tr}_A\left[\left(P_{A,F}^{p_A}\rho_{A,F}^{\sigma p_B}\right)+\left(P_{A,F}^{-p_A}\rho_{A,F}^{- \sigma p_B}\right)\right]^{n}\\
\fl\quad\quad \quad \quad=\frac{1}{2^{(n-1)(L-1)}} \text{Tr}_A\left[\frac{1+p_A \eta_A}{2}\text{Tr}_{B,F}\left[\ket{\phi(u)}\bra{\phi(u)} \frac{1+ \sigma p_B \eta_B}{2}\right]\right.\nonumber\\
+\left.\frac{1-p_A \eta_A}{2}\text{Tr}_{B,F}\left[\ket{\phi(u)}\bra{\phi(u)} \frac{1-\sigma p_B \eta_B}{2}\right]\right]^{n}\\
\fl\quad\quad \quad \quad=\frac{1}{2^{(n-1)(L-1)}} \text{Tr}_A\left[\text{Tr}_{B,F}\left[\ket{\phi(u)}\bra{\phi(u)} \right]\right]^{n}=\frac{1}{2^{(n-1)(L-1)}} \text{Tr}_A\left[\rho_{A,F}^{n}\right].
\end{eqnarray}

Finally, one notes that the state of a pure $Z_2$ gauge field $\ket{G(u)}$ is written as a gauge average of a given configuration $\ket{u}$ over all gauge-equivalent configurations $\ket{\tilde{u}}$, so that $\ket{G(u)}=\frac{1}{2^{(N+1)/2}}\sum_{\tilde{u}\sim u}\ket{\tilde{u}}$. If we compute $\text{Tr}_{A,G}\left[\rho^n_{A,G}\right]$ we then get
\begin{equation}
\text{Tr}_{A,G}\left[\rho^n_{A,G}\right]=\frac{1}{2^{(n-1)(L-1)}},
\end{equation}
 which thus means that 
\begin{equation}
\text{Tr}_A\left[\rho_A^n\right]= \text{Tr}_{A,G}\left[\rho^n_{A,G}\right]\text{Tr}_{A,F}\left[\rho^n_{A,F}\right].
\end{equation}
which is the desired result.

\end{document}